# Emergence of multiple relaxation processes during low to high density transition in $Au_{49}Cu_{26.9}Si_{16.3}Ag_{5.5}Pd_{2.3}$ metallic glass

Alberto Ronca[1,2*#], Antoine Cornet[1,2#], Jie Shen[1,2#], Thierry Deschamps[3], Eloi Pineda[4] Yuriy Chushkin[2], Federico Zontone[2], Mohamed Mezouar[2], Isabella Gallino[5], Gaston Garbarino[2] and Beatrice Ruta[1,3*]

[1] *Institut Néel, Université Grenoble Alpes and Centre National de la Recherche Scientifique, 25 rue des Martyrs - BP 166, 38042, Grenoble cedex 9 France*

[2] *European Synchrotron Radiation Facility, 71 avenue des Martyrs, CS 40220, Grenoble 38043, France*

[3] *Institut Lumière Matière UMR CNRS 5306, Université Claude Bernard Lyon 1, CNRS, F-69622 Villeurbanne, France*

[4] *Department of physics, Institute of Energy Technologies, Universitat Politècnica de Catalunya-BarcelonaTech, 08019 Barcelona, Spain*

[5] *Chair of Metallic Materials, Berlin Institute of Technology (TU-Berlin), Ernst-Reuter Platz 1, 10587 Berlin, Germany*

*Corresponding authors
alberto.ronca@neel.cnrs.fr
beatrice.ruta@neel.cnrs.fr

# These authors equally contributed

**Abstract**

The existence of multiple amorphous states, or polyamorphism, remains one of the most debated phenomena in disordered matter, particularly regarding its microscopic origin and impact on glassy dynamics. Profiting of the enhanced data quality provided by brilliant synchrotrons, we combined high pressure X-ray photon correlation spectroscopy and X-ray diffraction to investigate the atomic dynamics-structure relationship in a $Au_{49}Cu_{26.9}Si_{16.3}Ag_{5.5}Pd_{2.3}$ metallic glass at room temperature. We identify a structural and dynamical crossover near 3 GPa, marked by avalanches-like massive atomic rearrangements that promote the system toward increasingly compact atomic cluster connections. This crossover superimposes to a pressure-induced acceleration of the atomic motion recently reported, and signals the onset of a transitional state, potentially linked to the nucleation of a new phase within the glass, characterized by the coexistence of two amorphous states with distinct relaxation processes. These results provide evidence for a sluggish, continuous polyamorphic transformation, even in absence of marked structural discontinuities.

## 1. Introduction

The ability of matter to adopt multiple structurally distinct phases, known as polymorphism, is a hallmark of crystalline materials[1,2]. In the last decades, transitions between different amorphous states, termed polyamorphism, were also reported in several disordered systems[3,4]. Observations of liquid–liquid transitions (LLTs) and glass–glass transitions (GGTs) have revealed that amorphous materials can exhibit a richness of behaviours once thought to be exclusive to crystals. Yet, despite this growing body of evidence, the microscopic mechanisms that govern transitions between amorphous states, and their implications for the dynamics of glasses, remain poorly understood. Although GGTs are rare to found, their occurrence in different families of hard and soft glasses points toward their universal character. In soft materials, the transition between the two amorphous states is usually triggered by a modification of the effective interparticle interactions trough an external parameter, and has been reported for sticky hard-spheres colloids[5] and highly cross-linked microgel particles[6]. A spontaneous, temporal-driven GGT has been also detected in colloidal clays[7].

In structural glasses, hydrostatic pressure is generally the key thermodynamic variable driving transformations between distinct amorphous states. Polyamorphic transitions from a low-density amorphous state (LDA) to a high-density amorphous state (HDA) have been extensively documented in systems with directional bonding, such as amorphous ice[8], oxide glasses[9], chalcogenides[10] and silicon[11]. In all these cases, pressure induced a transformation from low-coordination local environments to densely packed configurations, independently of the specific nature of the atomic bonding. More recently, polyamorphism has also been observed in metallic glasses (MGs)[12–18], despite their densely packed atomic structure and predominantly non-directional bonding. In Ce-based MGs, the GGT has been attributed to changes in electronic interactions, leading to bond shortening and a concomitant volume collapse[19]. Differently, in Zr-based MGs, the polyamorphism is driven by locally favoured directional structures[20].

The possible connection between GGTs and thermodynamic first-order phase transitions has been the subject of longstanding debates in all systems[12,14,19,21–24]. A common counterargument arises from the observation that the LDA-HDA transformation is often not abrupt and covers a broad pressure range. This effect could be a simple consequence of the sluggish kinetics of glasses at low temperatures, where GGTs are usually measured[12].

A major obstacle for the understanding of polyamorphism in glasses is that most studies have focused exclusively on structural changes during GGTs, without considering the role played by the intrinsic relaxation processes characteristic of glassy states. Studies on aging in MGs have shown that even tiny, almost invisible structural transformations can produce changes in the atomic dynamics exceeding an order of magnitude[25–27]. It is therefore natural to expect pronounced changes in the particle motion accompanying polyamorphic transitions. Indeed, this has been observed in soft glasses, where GGTs have been directly associated to distinct particle dynamics[7,28].

In the case of MGs, X-ray Photon Correlation Spectroscopy (XPCS) studies have shown that the α-relaxation that governs the particle motion in supercooled liquids is frozen in the glass. Notwithstanding, MGs still exhibit atomic rearrangements at low temperatures whose dynamics is driven by intrinsic stresses stored in the material during the vitrification process[29]. The possibility to probe these dynamics under high pressure has become possible only recently thanks to the increase in coherent flux in modern synchrotron sources[30–32]. The first applications of high pressure XPCS in glasses have revealed a structure-dynamic decoupling during pressure-induced structural changes[32–34], even in the absence of polyamorphic transitions. In the case of a pre-annealed Ce-based MG, the density increases monotonously with pressure during a GGT while the dynamics is accelerated by the transition[33]. An acceleration of the dynamics has been observed also in hyper-quenched Pt-based MG during the free volume release, while further increasing the pressure induces dynamical arrest[34]. This non-monotonic behaviour has been linked to the formation of transient double local minima in the potential energy landscape (PEL) under pressure[35,36], which may also account for the observed rejuvenation and strain hardening reported in other MGs[37].

To elucidate the role of hydrostatic compression in governing the collective atomic motion in MGs and its interplay with pressure-induced structural changes, we performed a detailed investigation of the pressure dependence of the atomic motion and structure at room temperature in a hyper-quenched $Au_{49}Cu_{26.9}Si_{16.3}Ag_{5.5}Pd_{2.3}$ MG. This was achieved by combining high-pressure XPCS and X-ray Diffraction (XRD) synchrotron experiments. The selected alloy exhibits a relatively low glass transition temperature ($T_g = 396\ K$), excellent glass-forming ability (GFA), and a fragile character in the corresponding supercooled liquid, with a fragility index $m = 46.2$[38,39]. At ambient pressure, the $Au_{49}Cu_{26.9}Si_{16.3}Ag_{5.5}Pd_{2.3}$ exhibits a LLT[27,39] near $T_g$ which have been observed by quasi-static cooling from the supercooled liquid. The presence of such LLT suggests the possible existence of polyamorphism also in the solid glassy state, as previously identified in other compositions[40,41], including the analogous $Au_{55}Cu_{25}Si_{20}$ MG[17]. In concomitance to the structural changes associated to the LLT, the atomic dynamics of $Au_{49}Cu_{26.9}Si_{16.3}Ag_{5.5}Pd_{2.3}$ displays a pronounced fragile-to-strong crossover in the temperature dependence of the microscopic structural relaxation process[27], highlighting the strong coupling between dynamics and LLT.

Our XRD measurements reveal a subtle structural crossover near 3 GPa, consistent with a change in compressibility between two glassy states. Strikingly, this pressure value closely correlates with the emergence of an additional slow relaxation process in the dynamics, triggered by avalanche-like atomic rearrangements. We interpret these findings as evidence for the nucleation of a new amorphous state within the glassy matrix, marking the onset of a sluggish polyamorphic transition.

## 2. Experimental

*2.1 Materials preparation*

$Au_{49}Cu_{26.9}Si_{16.3}Ag_{5.5}Pd_{2.3}$ metallic glasses were produced by melting a mixture of high-purity elements (99.995%) at $\approx 1100$ K in an alumina crucible using an Indutherm MC15 casting apparatus. The melted alloy is then injected onto a rotating copper wheel under controlled argon atmosphere producing $\approx 20\ \mu m$ thick ribbons. Prior to all experiments, the amorphous nature of the material was confirmed by X-ray diffraction. The specimens were stored in a freezer at roughly 250 K to avoid low temperature physical aging.

*2.2 High pressure set-up for XRD and XPCS measurements*

The high-pressure setup has been described in detail in references[31,34]. A membrane-driven diamond anvil cell (DAC) loaded with nitrogen as pressure-transmitting medium (PTM) is used to apply pressure on the metallic glass, ensuring hydrostaticity in the studied range. The as-cast MG (with dimensions $\approx 100 \times 100 \times 20\ \mu m^3$) is inserted in a stainless-steel gasket with a 300 $\mu m$ diameter and 80 $\mu m$ thickness. A pressure by ruby luminescence (PRL) system is used to monitor the in-situ pressure. Pressure was found stable before and after each XPCS/XRD measurement within the experimental uncertainty of 0.1 GPa.

*2.3 High pressure XRD*

The structure of the metallic glass sample under in-situ high pressure was monitored using XRD at the ID27 beamline of the European Synchrotron (ESRF, Grenoble, France), with an incident beam energy of 33 keV. Scattered intensity was recorded with a Eiger2 X CdTe 9M (active area = 233.1 x 244.7 mm$^2$, pixel size = 75 µm) with a sample-detector distance of 0.248 m allowing to probe a maximum scattering vector of 12 Å$^{-1}$. Azimuthal integration of 2D diffraction patterns was performed using routines from the pyFAI library[42,43] to compute 1D diffraction patters. The python-based Amorpheus software[44] was then used to calculate the (background corrected) Faber-Zimman structure factor $S(q)$ with Krogh-Moe-Norman normalization[45]. To account for possible asymmetries, we modelled the top of FSDP using an ad-hoc function $S(q) = y_0 + A\left[1 + e^{-\frac{q-q_c+\omega_1/2}{\omega_2}}\right]^{-1} \times \left[1 - \left(1 + e^{-\frac{q-q_c-\omega_1/2}{\omega_3}}\right)^{-1}\right]$. The top 60%, 50%, 40% and 30% of the FSDP was fitted to estimate the center of mass and the width of the reconstructed peak. The 50% threshold was chosen arbitrarily in this work and the consistency of the results when using a different portion of the peak was verified. The $S(q)$ is Fourier-transformed to calculate the reduced pair distribution function $G(r)$ as $G(r) = 4\pi r[(\rho(r) - \rho_0] = \frac{2}{\pi}\int q(S(q) - 1)\sin(qr)\,dq$, where $\rho(r)$ is the atomic pair distribution function and $\rho_0$ is the average atomic density. Being the investigated $Au_{49}Cu_{26.9}Si_{16.3}Ag_{5.5}Pd_{2.3}$ MG a five-component system, the total $G(r)$ comprises all 15 partial pair distribution functions of each atomic pair, weighted by the X-ray scattering weight $w_{ij}$[46] which directly depends on the atomic fraction of each constituent elements. The different shells of the $G(r)$ are then fitted by using a single or a sum of Gaussian functions.

*2.4 High pressure XPCS*

XPCS experiments were performed at ID10 beamline at ESRF (Grenoble, France), using an incident beam energy of 21.67 keV (ΔE/E=1.4×10$^{-4}$). The beam was collimated by high-power slits and focused using 2D Be lenses to a beamsize of 5.2x4.2 µm$^2$ (HxV, FWHM) on the sample. Guard slits were used to minimize incoherent background from stray radiation around the beam. The measured photon flux was 7.3x10$^{11}$ photon/s. Speckle patterns were collected in a wide-angle geometry with an Eiger2 4M CdTe detector placed 5 m downstream from the sample position with an acquisition time of 0.1 or 0.5s per frame. The dynamics was measured at the maximum of the FSDP, $q = 2.68$ Å$^{-1}$ in standard condition, by changing the scattering angle to match the pressure evolution of the FSDP. The sparse correlator algorithm[47] was used to extract the two-times correlation functions (TTCFs) defined as $G(q,t_1,t_2) = \frac{\langle I(q,t_1) \times I(q,t_2) \rangle}{\langle I(q,t_1) \rangle \times \langle I(q,t_2) \rangle}$, where the average symbol represents the ensemble average over all detector pixels. This quantity provides the instantaneous correlation of the scattered intensity measured at two subsequent time $t_1$ and $t_2$ as further detailed in ref[48]. The $g_2(q,t)$ intensity-intensity correlation functions are then extracted from the TTCF by averaging over different time intervals. The Siegert relation $g_2(q,t) = 1 + \gamma |F(q,t)|^2$, $\gamma$ being the experimental contrast, is used to estimate the intermediate scattering function $F(q,t)$ from the $g_2(q,t)$. The application of the Siegert equation to non-ergodic systems like glasses is ensured by the many q-equivalent speckles in large area detector. Decay functions are then modelled using the KWW equation. The temporal dependence study at 3.2 GPa is performed by time-averaging portion of the TTCF to compute ISFs at different waiting times $t_w$ and over a defined temporal window. $t_w$ is defined from the time elapsed from the pressure change as $(t_1 + t_2)/2$, $t_1$ and $t_2$ being the starting and ending time of the XPCS measurement. Each ISF is then fitted with the KWW model to extract $\tau$ and $\beta$ corresponding to the given $t_w$. Kossel lines from the diamond are masked from the scattering pattern before applying the sparse correlator algorithm with the DAC contributing only statically to the observed ISFs. Approximately two hours were spent at each pressure point, during which several XPCS scans were acquired (Fig. S7).

### 3. Results

*3.1 Pressure evolution of the structure*

To establish a quantitative connection between structural changes measured by XRD and atomic dynamics probed by XPCS, we selected a pressure range of 0 – 7.5 GPa. This range satisfies the stability requirements of XPCS, ensures hydrostatic conditions, and corresponds to the regime where the onset of polyamorphism is expected in many MGs[12,16]. Figure 1 shows the first two diffraction peaks in the static structure factor, $S(q)$, measured during compression, along with the fitting parameters describing the first sharp diffraction peak (FSDP). The sample remains fully amorphous during the entire pressure cycle. Consistent with structural densification under compression, the position of the maximum of the FSDP, $q_1$, shifts toward higher $q$-values with pressure, while its intensity increases (insets of Fig. 1a). Both $q_1$ and the maximum of the second diffraction peak, $q_2$, follow similar linear relations with pressure, $5.15 \times 10^{-3}$ Å$^{-1}$/GPa for $q_1$ (Fig. 1b) and $1.1 \times 10^{-2}$ Å$^{-1}$/GPa for $q_2$ (Fig. S1). The FSDP position, $q_1$, can be used as an indication of density changes in the glass[49,50]. This correlation is supported in our composition by the strong agreement between the thermal expansion coefficient estimated via dilatometry[51] and the relative change of $q_1$ obtained by XRD[51]. Accordingly, compression induces a monotonic reversible increase in density, with no apparent permanent structural changes after decompression (Fig. 1b), consistent with previous observations in $Cu_{46}Zr_{46}Al_8$[52] and $Pt_{42.5}Cu_{27}Ni_{9.5}P_{21}$[34] MGs. In $Au_{49}Cu_{26.9}Si_{16.3}Ag_{5.5}Pd_{2.3}$, hydrostatic compression up to 7.5 GPa corresponds to an estimated density increase of 3.7 – 4.4%, depending on the model employed to calculate the volumetric changes (see Fig. S2 and related discussion). By comparison, other MGs under similar loading conditions exhibit density increments ranging from 5% to 20%, depending on the material's bulk modulus[53].

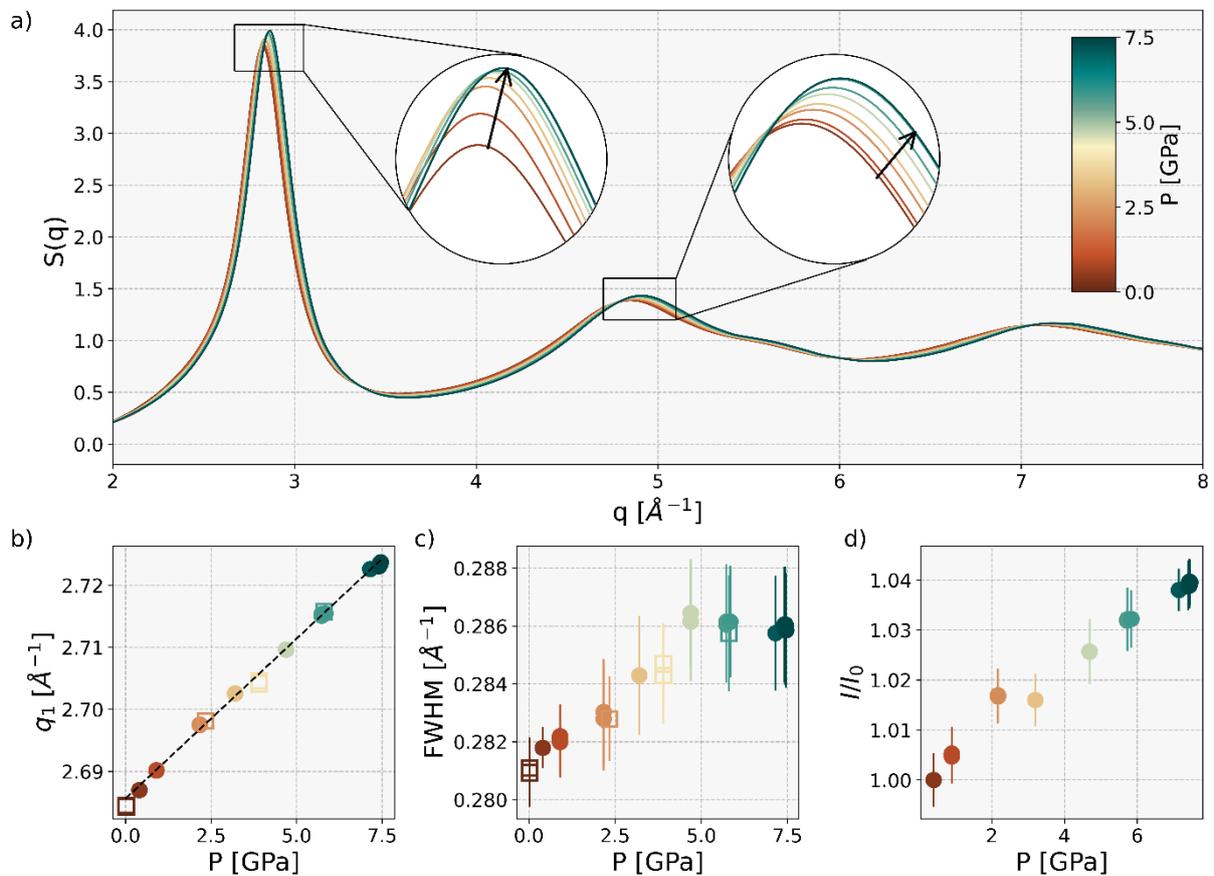

**Fig. 1. Pressure dependence of the static structure factor. a)** Portion of the $S(q)$ measured by XRD during compression at different pressures. The insets show the pressure evolution of the top of the first two diffraction peaks. Pressure dependence of the FSDP: peak position **(b)**, full width at half maximum (FWHM) **(c)** and normalized intensity **(d)**. Data acquired on compression and decompression are shown as filled circles and empty squares, respectively.

Differently from $q_1$, the full width at half maximum (FWHM) and the intensity, $I(q_1)$, of the FSDP exhibit a less trivial behaviour. While the FWHM reaches a plateau at around ≈ 4 GPa following a first initial linear increase of around 2%, $I(q_1)$ shows two distinct regimes, with a noticeably reduced pressure dependence beyond ≈ 3 GPa (Fig. 1c and 1d, and Fig. S3 and S4, respectively). In the 2 – 3 GPa range, $I(q_1)$ is nearly constant, marking the transition between the initial rapid increase and the subsequent slower growth at higher pressure.

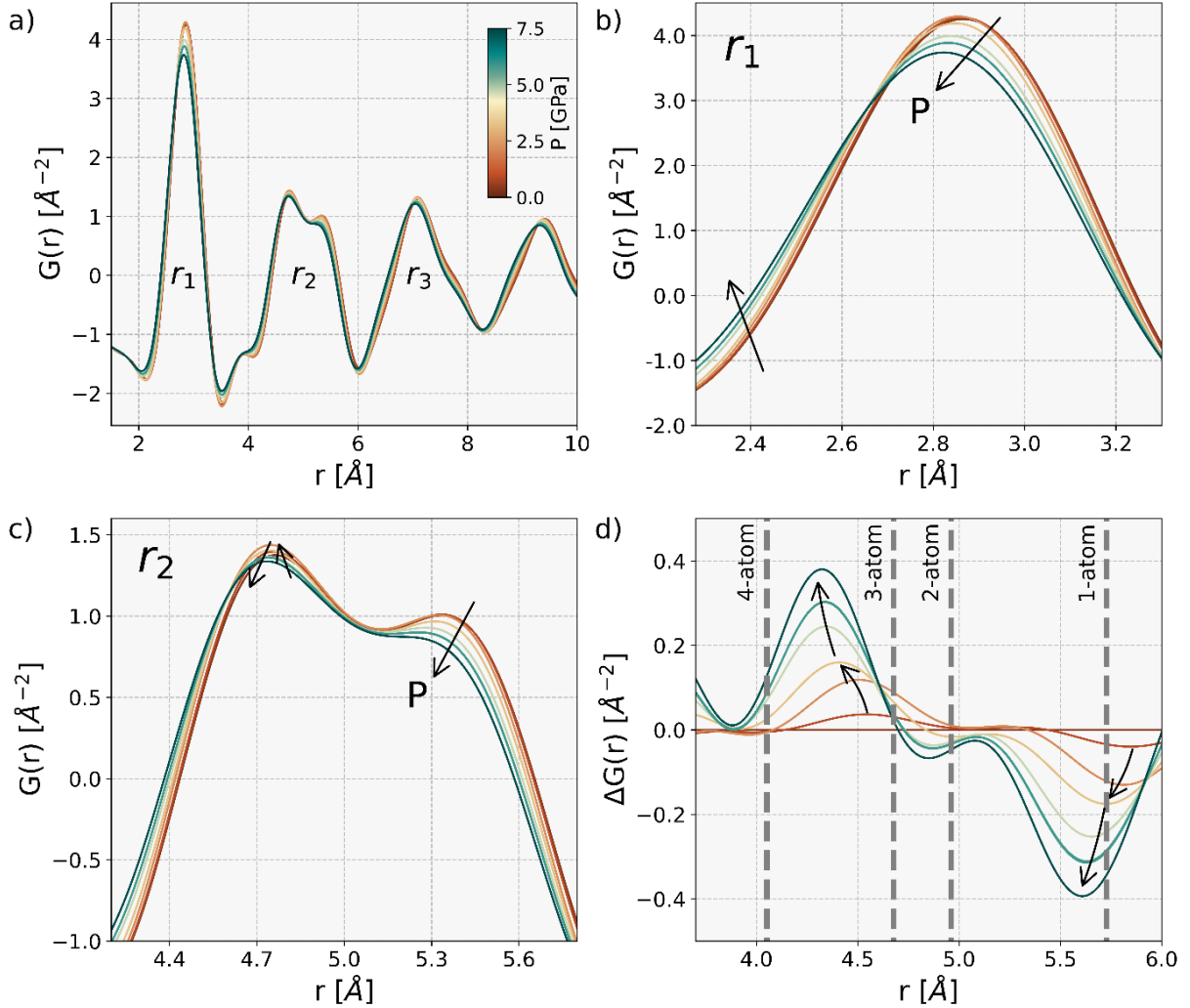

**Fig. 2. Pressure dependence of the reduced pair distribution function. a)** $G(r)$ evolution during compression. Magnification of the 1$^{st}$ **(b)** and 2$^{nd}$ **(c)** coordination shells (see Fig. S6 for the third shell). **d)** Differential pair distribution function $\Delta G(r, P) = G(r, P) - G(r, P = 0.4\ GPa)$ during compression. Vertical lines indicate the most probable second neighbour distances for cluster sharing $1 - 4$ atoms. Arrows mark the pressure behaviour of each peak.

To get additional insights into these subtle structural changes, we calculated the reduced pair distribution function $G(r)$ through a sine Fourier transformation of the $S(q)$ (Fig. 2). Table S1 summarizes the predominant atomic pairs responsible for the shape of the $G(r)$, mainly the Au-Au and the Au-Cu pairs, whose estimated interatomic distances at ambient conditions are ≈ 2.88 Å and ≈ 2.72 Å, respectively. The positions of each shell clearly shift towards lower values, confirming the densification under pressure. The first maximum of the $G(r)$ exhibits a decrease in the peak intensity which is accompanied by the emergence of a left shoulder with pressure (as indicated by the arrows in Fig. 2b). This behaviour can be interpreted as a decrease in the number of dominant Au-Au pairs, and a subsequent increase in the population of Au-Cu and potentially Au-Si and Cu-Cu pairs, with narrower bond length ($\approx 2.5 - 2.6$ Å at ambient conditions). The second peak of the $G(r)$ (Fig. 2c) shows the typical splitting observed in different MGs, with the two peaks exhibiting different intensity evolutions with pressure. The origin of this splitting is related to the presence of multiple cluster connection schemes[54,55]. In this view, neighbouring atomic clusters may share 1 to 4 atoms and the characteristic distances of each connection scheme can be approximated as: $2r_1$, $\sqrt{3}r_1$, $\sqrt{8/3}\,r_1$ and $\sqrt{2}r_1$ for 1-atom, 2-atoms, 3- and 4-atoms connectivity respectively, $r_1$ being the nearest-neighbour distance[54–56]. The evolution of these local structural motifs with pressure is presented in Fig. 2d in terms of the differential function $\Delta G(r, P) = G(r, P) - G(r, P = 0.4\ GPa)$. We observe a decrease in the 1-atom contribution with increasing pressure, accompanied by a corresponding rise in intensity at intermediate distances between the expected 3- and 4-atom connection scheme positions. Interestingly, both positions exhibit a change above approximatively 3 GPa, among two different trends, as indicated by the different arrows in the figure. The evolution of these structural motifs indicates a progressive reorganization of the local environment under compression, favouring more compact and topologically constrained structures over looser configurations. This tendency toward denser, more constrained local structures is further enhanced above $\approx$ 3 GPa.

To get quantitative information, Fig. 3 displays the fitting parameters obtained by modelling the first shell of the $G(r)$ using a single Gaussian function, and the second with a sum of two Gaussians. $r_1$ corresponds to the first shell, while $r_{21}$ and $r_{22}$ refer to the first and second peak of the second shell, respectively (see Fig. S5 for fitting examples). Whereas the maximum position of $r_1$ exhibits a unique linear relation, two distinct linear regimes emerge for $r_{21}$ and $r_{22}$, with a mild but noticeable change in slope occurring at $\approx$ 3 GPa, above which the dependence on pressure weakens. Evident changes in the pressure dependencies of the FWHM and intensities of all peaks can be observed around the same pressure value. Beyond this threshold, the system evolves into a different structure, as evidenced by the reduction in peak intensity and the broadening of the FHWM of $r_1$. The pressure evolution of the $r_{21}$ intensity matches the one observed for the FSDP, with two regimes manifesting below and above $\approx$ 3 GPa, while the intensity of $r_{22}$ is decreasing almost linearly in the probed pressure range. The $r_{21}$ FHWM instead broadens above ≈ 3 GPa, similarly to what observed for $r_1$, whereas the $r_{22}$ FHWM is slightly varying with pressure. Despite the small variation (0.02, 0.05 and 0.01 for $r_1$, $r_{21}$ and $r_{22}$, respectively), a clear inversion of trend can be detected in the pressure dependence of FWHM between 2 and 3 GPa, consistently with what is observed in the other pressure-dependencies of the fitting parameters, suggesting the existence of a structural crossover in the response of the system at all length scales (see also the third shell $r_3$ of the $G(r)$ in Fig. S6).

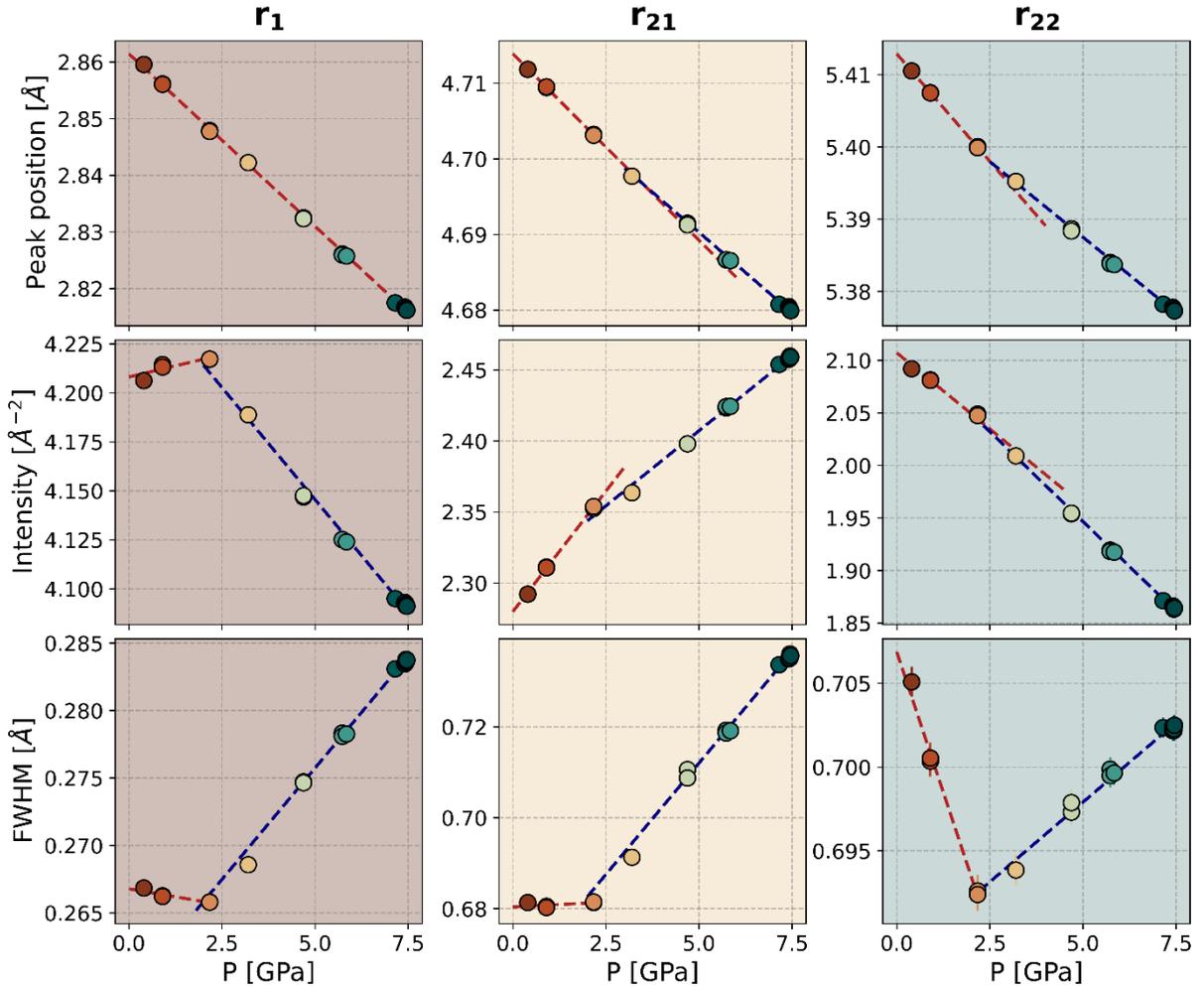

**Fig. 3. Pressure dependence of the first ($r_1$) and second coordination shells ($r_{21}$ and $r_{22}$) of the $G(r)$.** Peak positions (**top row**), intensities (**middle row**) and FWHM (**bottom row**). Dotted lines are linear fit of the data. A single linear fit fails to capture the pressure evolution of the peak positions of the second shell ($r_{21}$ and $r_{22}$); instead, two distinct linear regimes – below and above ≈ 3 GPa – yields an improved representation of the data. All the other fitting parameters exhibits a more or less evident change in the 2 – 3 GPa pressure range, and are modelled accordingly (red and blue dotted lines).

*3.2 Pressure evolution of the dynamics*

It is well known that weak structural changes, can be accompanied by important dynamical contributions[6,7,26]. This effect is a simple consequence of the higher sensitivity of second order correlation functions with respect to static observables. To verify the presence of a possible crossover in the dynamics at $\approx 3$ GPa, XPCS data were collected in the correspondence of the FSDP ($q = 2.68$ Å$^{-1}$ at 1 atm and room temperature) during isobars in the pressure range of $0.9 - 7.5$ GPa and for different waiting times $t_w$, estimated as the elapsed time between the data acquisition and each pressure change. Details on the experimental protocol and data analysis can be found in the *Material and Methods* section.

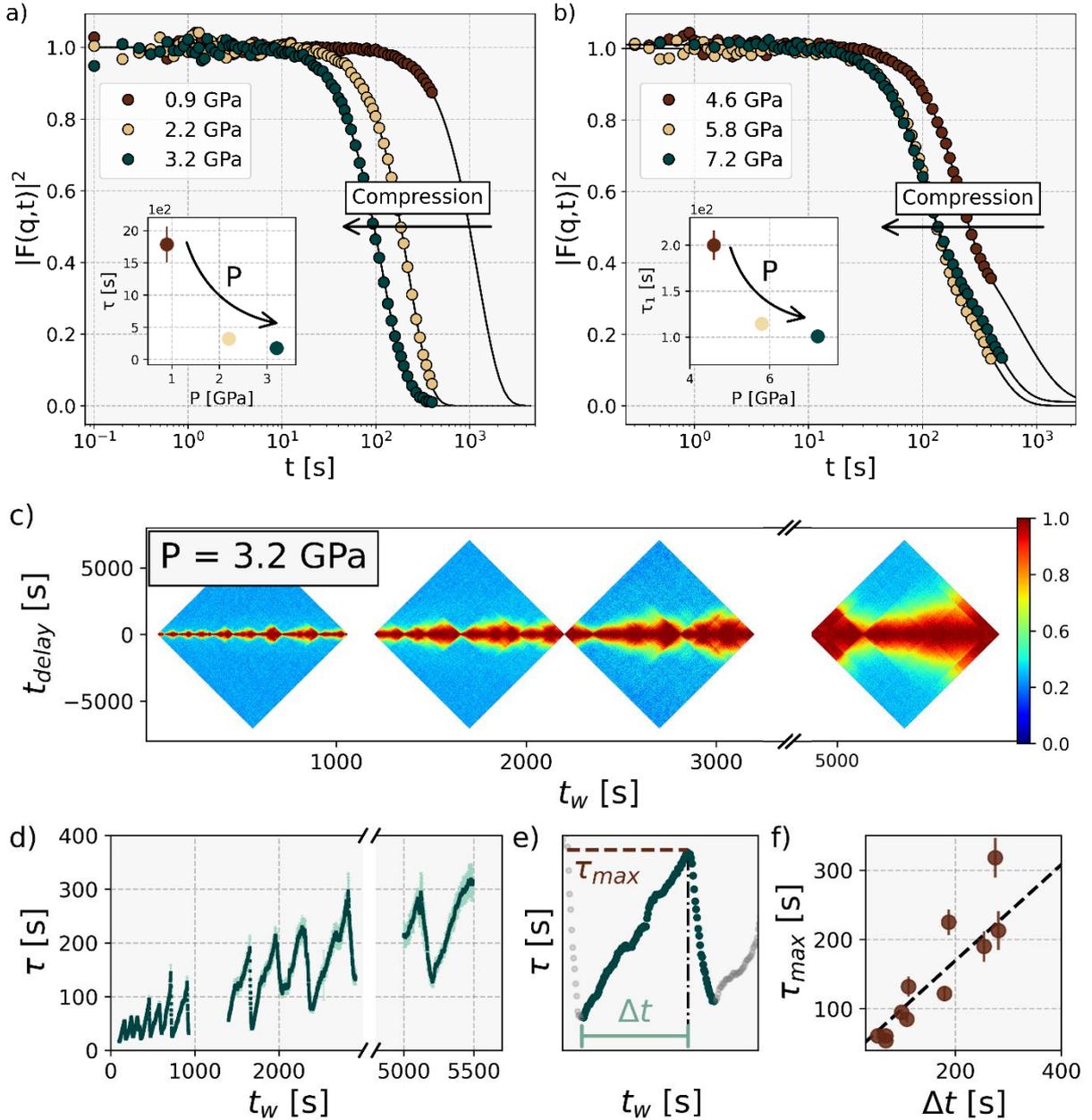

**Fig. 4. Pressure dependence of the atomic dynamics.** Selection of ISFs measured at different pressures during compression below **(a)** and above **(b)** 3.2 GPa. Data are evaluated at similar waiting times from pressure changes to allow for an effective comparison (see also Fig. S8 and S10 for full decorrelation). Black solid lines are KWW fits of the data. The corresponding characteristic times of the main relaxation process are displayed in the insets. Error bars are within the size of the data points when not showed. **c)** Successive TTCFs acquired at 3.2 GPa as a function of the waiting time $t_w$, showing heterogeneous and avalanche-like dynamics. **(d)** Temporal evolution of the structural relaxation time at 3.2 GPa. **(e)** Scheme of a typical decorrelation event with $\tau_{max}$ indicating the slowest relaxation time reached before the avalanche, and $\Delta t$ the temporal interval prior to a successive massive atomic rearrangement. **(f)** Evolution of $\tau_{max}$ plotted against the time span $\Delta t$ between two consecutives avalanches. The dotted line is a linear fit.

The atomic dynamics is described by the intermediate scattering function (ISF) $F(q,t)$, which monitors the temporal decay of the density-density correlation functions at the given $q$ and pressure conditions. The normalised intensity auto-correlation functions measured with XPCS, $g_2(q,t)_{norm}$, provide information on the ISF and are modelled using the empirical Kohlrausch Williams Watt (KWW) function: $g_2(q,t)_{norm} = |F(q,t)|^2 = \exp(-2(t/\tau)^\beta)$, where $\tau$ is the relaxation time, representing the characteristic time of the atomic rearrangements and $\beta >1$ is the compressed shape parameter, signature of a stress-driven dynamics in the glass state[57–59]. Fig. 4 shows a selection of auto-correlation functions measured during compression. To compare different pressures, the data have been evaluated for similar waiting times from the pressure onset. For this reason, some of the curves are incomplete at longer times (see Fig. S8 and S10 to observe the full decay at each pressure). As previously observed in another as-cast MG[34], although the structural densification, the atomic dynamics dramatically accelerates during loading at early compression stages (Fig. 4a). This acceleration is evidenced by a shift in the ISF of more than one order of magnitude towards faster time scales on increasing pressure. In agreement with previous theoretical works[35,36], this dynamical acceleration can be associate to the pressure-induced emergence of a second metastable state in the PEL at energies higher than that of the initial minimum, allowing fast transitions between the two local minima[35,36]. In the present case, the acceleration persists in the whole pressure range (Fig. 4b), despite showing a milder decrease of $\tau$ with increasing pressure: from a $\Delta\tau \approx -90\%$ between 0.9 and 3.2 GPa, to a $\Delta\tau \approx -50\%$ from 4.6 to 7.2 GPa, as can be seen in the $\tau(P)$ dependences (insets of Fig. 4a and 4b). At each pressure (but 3.2 GPa which is discussed below), the dynamics is homogenous and accompanied by continuous aging (Fig. S8a and S8b).

For pressures above 3.2 GPa, a second decay emerges in the ISFs, superimposed on the predominant compressed decorrelation component. This additional relaxation occurs at time scales almost one order of magnitude longer and exhibits reduced strength relative to the main process (Fig. 4b and S8). The double-decay behaviour persists when probing a fresh, previously unexposed region of the sample, as shown in Fig. S9. The distinct atomic dynamics in the two pressure regimes are directly compared in Fig. S8. Remarkably, the onset of this slow relaxation process coincides with the pressure range of the structural crossover in the XRD data, indicating a strong correlation between atomic structure and dynamics (Fig. 1 and 3). We note that a similar evolution of the ISFs has been observed during the high- to low-density polyamorphic transition in amorphous ice[60–62], where it accompanies important structural changes. Differently, no experimental evidence of such behaviour has been reported previously in MGs.

The modelling of two-steps ISFs requires a sum of two KWW terms: $|F(q,t)|^2 = \left[a \exp\left(-\left(\frac{t}{\tau_1}\right)^{\beta_1}\right) + (a-1) \exp\left(-\left(\frac{t}{\tau_2}\right)^{\beta_2}\right)\right]^2$, $a$ and $(1-a)$ being the amplitudes of the two relaxation modes (see Fig. S8 and S10 for further details). Here, $(\tau_1, \beta_1)$ and $(\tau_2, \beta_2)$ represent the relaxation times and shape parameter of the primary and secondary relaxation processes, respectively. To compare the effect of pressure on the dynamics, the primary relaxation time $\tau_1$ is considered, as showed in the inset of Fig. 4b. Both relaxation processes can be described by compressed exponential decays. This means that although the different strength and characteristic time, both processes result from the stress-driven anomalous diffusion[58,59] usually observed in MGs at atomic scale.

The transition between the two dynamical regimes occurs at 3.2 GPa and is accompanied by the surprising activation of pronounced avalanche-like atomic rearrangements, persisting for at least ~2h, corresponding to the duration of our isobaric steps. These abrupt collective motions results in a highly heterogeneous evolution in the two-times correlation functions (TTCFs) (Fig. 4c), which provide a time-resolved representation of the ISFs, with the width of the high-intensity diagonal contour proportional to $\tau$ (see *Material and Methods*). The temporal evolution of $\tau$ extracted from the TTCFs is reported in Fig. 4d. The data reveal nearly instantaneous decorrelations within the scattering volume, manifested as sudden drops in $\tau$, followed by progressively slower dynamics as the elapsed time increases.

These heterogeneous dynamics can be characterized by two representative parameters, as schematized in Fig. 4e: the maximum relaxation time reached prior to an avalanche ($\tau_{max}$), and the temporal interval between subsequent avalanches ($\Delta t$). Fig. 4f shows the evolution of $\tau_{max}$ as a function of $\Delta t$, which is well modelled with a linear equation, revealing a direct correlation between the occurrence of large-scale collective rearrangements and physical aging.

This correlation indicates that the frequency of avalanches decreases as the stability of the probed configuration - and thus its relaxation time - increases. As detailed in the SI (see Fig. S11), we ruled out any possible artefact arising from the set-up or the pressure-transmitting medium. Notably, the extraordinary coexistence of both physical aging[26] and intermittent-like avalanche dynamics[63,64] at 3.2 GPa points to a transitional state in the MG, consistent with the observed structural changes in both S(q) and G(r) and the emergence of a second relaxation process in the ISFs at larger pressures.

## 4. Discussion

The XPCS and XRD measurements reveal a non-trivial dynamical response of the $Au_{49}Cu_{26.9}Si_{16.3}Ag_{5.5}Pd_{2.3}$ MG to pressure-induced structural modifications. While increasing pressure promotes progressively more compact atomic cluster-cluster connections (Fig. 3d) and a density increase of 3.7 – 4.4% (Fig. S2), the atomic motion accelerates dramatically, by more than one order of magnitude across the investigated pressure range. This acceleration is consistent with previous investigations[34] and theoretical predictions[35,36] of the pressure-promoted emergence of transient secondary high-energy metastable states within the PEL. A clear deviation in the pressure dependences of the structural observables emerges in the 2 – 3 GPa pressure range, as evidenced by the existence of a transition in the evolution of the peak positions, intensities and widths of the FSDP and the first, second and third coordination shells of the $G(r)$ (Fig. 1 – 3 and Fig. S6).

It is important to stress that the pressure evolution of the peak positions ($q_1$, $r_1$ and $r_{12}$) in the structural data does not suggest a first order transition; rather, the associated volume contraction and expansion can be accurately described using a single equation of state (Fig. S2). On the contrary, the existence of a transition between two different states is supported by the pressure derivative of the volume – i.e., a measure of the compressibility[65] – which reveals two distinct regimes: an initial sharp decrease followed by a plateau beyond the ≈ 3 GPa threshold (Fig. 5). This pressure response is confirmed by employing an alternative method for the estimation of the compressibility (Fig. S2 and related discussion). While the observed decrease in compressibility in the first regime is reported also in other metallic glasses[66,67] and is a consequence of the reduced ability to shrink high-density MGs upon compression, the emergence of a stationary regime at larger pressures indicates a lower sensitivity to pressure at higher loads, in line with what is observed for the FWHM and $I(q)$ of the FSDP which are representative of the medium range order (MRO)[20] (Fig. 1c and 1d) and with the weaker pressure dependence of the relaxation time above 3.2 GPa (Fig. 4b). These findings align with the notion of a saturation in the pressure response of the elastic properties of MGs at high pressures[66]. It is worth noticing that similar sudden changes in compressibility have been observed in MGs after the occurrence of polyamorphic transitions[68,69] between a low- to high-density phase, strengthening the idea of the occurrence of a similar process also in the probed composition.

While structural changes can be subtle and difficult to interpret uniquely, the atomic dynamics also shows dramatic changes in the proximity of the crossover pressure, confirming the existence of a transition as reported by the atomic structure. At 3.2 GPa, the TTCFs indicates severe atomic rearrangements (Fig. 4c), showing the co-existence of physical aging and avalanche-like massive atomic displacements whose frequency decreases for larger relaxation times and thus more stable atomic configurations. At larger pressures, the avalanches disappear, the dynamics returns homogeneous but an additional relaxation process emerges in the ISFs at almost one order of magnitude longer time-scales and with reduced strength with respect to the main relaxation component (Fig. 4b, Fig. S8). This secondary process exhibits a compressed exponential form of the ISF implying that it also originates from a stress-driven dynamics as usually observed in MGs at the atomic level[57–59]. Importantly, the same behaviour is observed when probing different, previously unirradiated sample regions, suggesting its origin as a bulk feature rather than a localized irradiation effect (Fig. S9). We note that while the emergence of pressure-induced multiple relaxation processes has been never observed in MGS, two-steps decays in the ISFs have been reported during the temperature-driven high- to low-density polyamorphic transition in powder amorphous ice[60]. In this case, however, the secondary dynamic component emerges at faster time-scales being associated to the low-density state and has been attributed to the coexistence of the two amorphous states. An additional oscillatory component was also reported in the ISFs of free-standing amorphous ice films[62], and its origin was attributed to the coexistence of two dynamically distinct components, consisting of lower density static domains within a higher density liquid-like matrix. In this context, the pronounced atomic rearrangements observed at 3.2 GPa in the Au-based MG (Fig. 4c-e), along with the consequent emergence of a slower secondary relaxation process at higher pressures, suggest the nucleation of a distinct amorphous state within the pre-existing glassy matrix – ultimately leading to the coexistence of two amorphous states during the densification process. Our interpretation is depicted in Fig. 5 where it is connected to the pressure-dependence of the system's compressibility. Unlike the case of amorphous ice, where the two phases are structurally distinct, the transition observed here is characterized by a very subtle structural changes, accompanied by a shift in the preferred atomic connection scheme and a reduction of the predominant Au-Au atomic pairs, followed by an increased population of the narrower Au-Cu, Au-Si and Cu-Cu pairs (Fig. 2b). These weak changes agree with those reported in other polyamorphic transitions in MGs, where GGTs often manifest as discontinuities or slope variations in the pressure-dependence of the first and second peak positions in the $S(q)$, with a continuous nature, spanning a pressure range of several GPa[12,18,19]. Accordingly, our findings suggest the onset of a sluggish polyamorphic transition, revealed by minor but still evident structural changes and, more clearly, by the highly sensitive dynamical observables. The different relaxation dynamics observed in the Au-based and Ce-based MGs in proximity of a polyamorphic transition may suggest distinct mechanisms responsible for such transformation. In the low density rare earth Ce-based MG, changes in the electronic structure drives the system into a different amorphous state[19,33]. Instead, the highly dense Au-based MG may undergo configurational polyamorphism, characterized by changes in bonding length and orientation, similar to the ones occurring in covalent-like MGs[8–11].

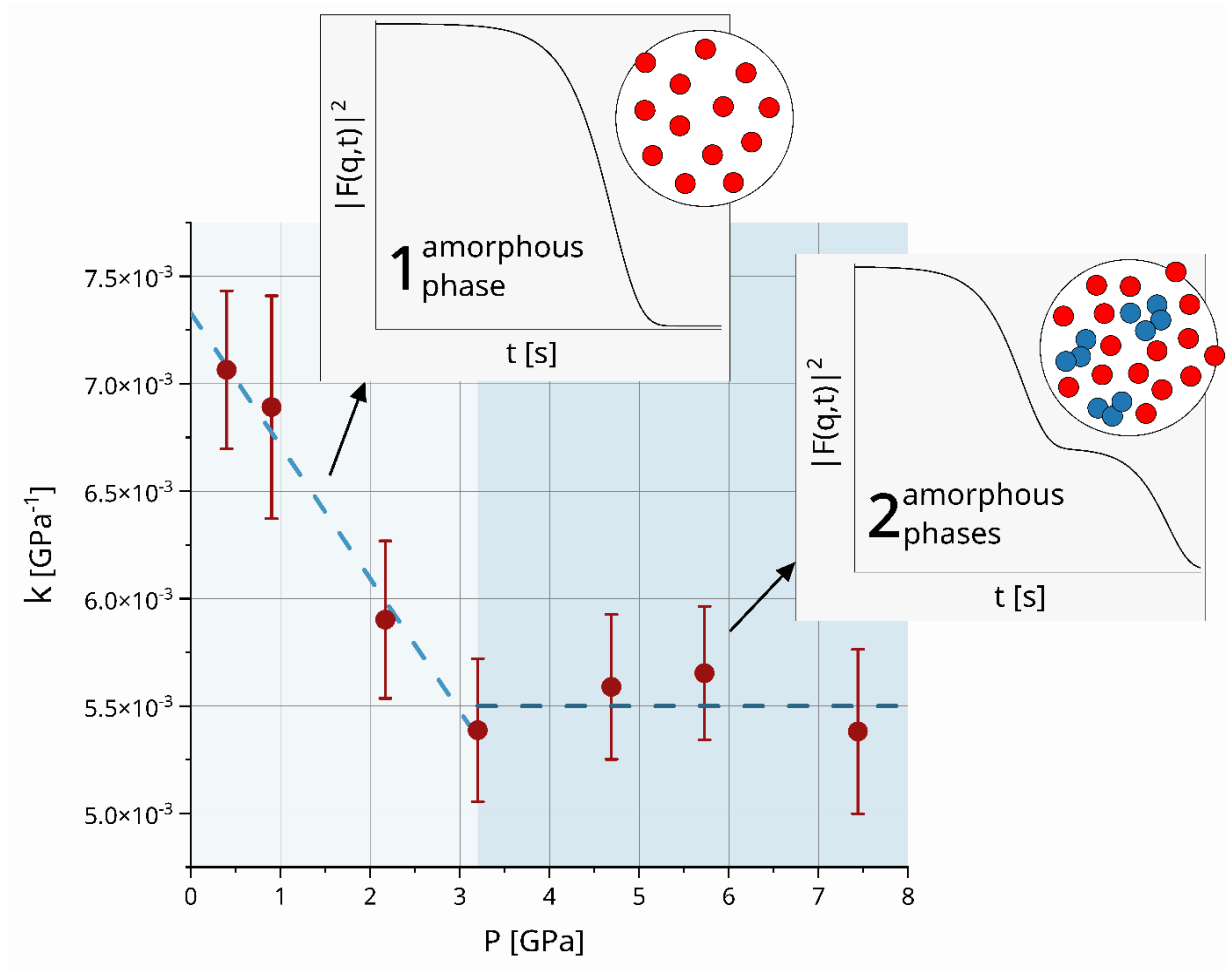

**Fig. 5. Pressure dependence of compressibility and corresponding atomic dynamics.** Isothermal compressibility, estimated from the numerical pressure derivative of the macroscopic volume, exhibiting two distinct regimes in pressure: an initial sharp decrease up to ≈ 3 GPa (light blue region), followed by a plateau at higher pressures (dark blue region). In the low-pressure region, the dynamics displays a single relaxation process in the ISF representing the initial amorphous state (red clusters in the circular inset). Beyond 3.2 GPa, a second, slower relaxation component emerges in the ISF, indicative of the occurrence of two distinct relaxation process within the sample. This dynamical behaviour is interpreted as the emergence and coexistence of two amorphous states within the denser glassy matrix, schematically represented by red and blue clusters.

It is important to stress that both the structural and dynamical crossovers were independently observed at the same pressure ( ≈ 3 GPa) employing two complementary synchrotron techniques, XRD and XPCS, respectively. The alignment of these two distinct measurements, performed in separate experiments, provides evidence of a real transformation occurring in the system.

## 5. Conclusions

In conclusion, we demonstrate that the $Au_{49}Cu_{26.9}Si_{16.3}Ag_{5.5}Pd_{2.3}$ MG undergoes a pressure-induced crossover in in the proximity of 3 GPa, with changes in both atomic structure and dynamics. While the atomic structure evolves toward higher density, a drastic change of compressibility appears at the crossover, where the microscopic dynamics exhibit massive avalanche-like rearrangements. At higher pressure, a slower secondary relaxation process emerges in the correlation function, consistent with the nucleation of a second amorphous state. These results indicate that pressure induces a sluggish polyamorphic transformation, leading to a denser glassy matrix in which the two amorphous states coexist. Our combined structural-dynamical approach provides a sensitive tool to uncover hidden transformations, such as polyamorphism or phase coexistence, in disordered materials under extreme conditions.

**Declaration of Competing Interest**

The authors declare that there are no financial or non-financial competing interests.

**Acknowledgements**

We gratefully acknowledge the ESRF (Grenoble, France) for providing beamtime, including experiments carried out on ID10 and ID27 beamlines under the LTP project HC4529. The Partnership for Soft Condensed Matter (PSCM) at the ESRF is also acknowledged for providing support facilities for sample characterization. We thank K. Lhoste and S. Bauchau for assistance on the ID10 and ID27 beamlines, and J. Jacobs for help with the high-pressure experiments. This project received funding from the European Research Council (ERC) under the European Union's Horizon 2020 research and innovation program (Grant Agreement No 948780).

**Data availability statement**

The data are available from the corresponding authors upon reasonable request.

**Author Contributions**

AC, JS, AR, EP, TD, YC, FZ and BR conducted the XPCS experiment. AC, JS, and BR carried out the XRD experiments. GG and MM assisted the HP-XPCS measurements. AR analyzed the XRD data, and AR and BR analyzed the XPCS data. TD performed the Raman spectroscopy experiments. AR and BR wrote the manuscript with contribution from all authors. BR conceived the work and led the investigation.

**References**

[1] J. Bernstein, Polymorphism – A Perspective, Crystal Growth & Design 11 (2011) 632–650. https://doi.org/10.1021/cg1013335.
[2] A.J. Cruz-Cabeza, S.M. Reutzel-Edens, J. Bernstein, Facts and fictions about polymorphism, Chem. Soc. Rev. 44 (2015) 8619–8635. https://doi.org/10.1039/C5CS00227C.
[3] H. Tanaka, Liquid–liquid transition and polyamorphism, The Journal of Chemical Physics 153 (2020) 130901. https://doi.org/10.1063/5.0021045.
[4] O. Mishima, Y. Suzuki, Propagation of the polyamorphic transition of ice and the liquid–liquid critical point, Nature 419 (2002) 599–603. https://doi.org/10.1038/nature01106.
[5] K.N. Pham, A.M. Puertas, J. Bergenholtz, S.U. Egelhaaf, A. Moussaïd, P.N. Pusey, A.B. Schofield, M.E. Cates, M. Fuchs, W.C.K. Poon, Multiple Glassy States in a Simple Model System, Science 296 (2002) 104–106. https://doi.org/10.1126/science.1068238.
[6] T. Eckert, E. Bartsch, Re-entrant Glass Transition in a Colloid-Polymer Mixture with Depletion Attractions, Phys. Rev. Lett. 89 (2002) 125701. https://doi.org/10.1103/PhysRevLett.89.125701.
[7] R. Angelini, E. Zaccarelli, F.A. De Melo Marques, M. Sztucki, A. Fluerasu, G. Ruocco, B. Ruzicka, Glass–glass transition during aging of a colloidal clay, Nat Commun 5 (2014) 4049. https://doi.org/10.1038/ncomms5049.
[8] C.A. Tulk, R. Hart, D.D. Klug, C.J. Benmore, J. Neuefeind, Adding a Length Scale to the Polyamorphic Ice Debate, Phys. Rev. Lett. 97 (2006) 115503. https://doi.org/10.1103/PhysRevLett.97.115503.
[9] J.P. Itie, A. Polian, G. Calas, J. Petiau, A. Fontaine, H. Tolentino, Pressure-induced coordination changes in crystalline and vitreous GeO 2, Phys. Rev. Lett. 63 (1989) 398–401. https://doi.org/10.1103/PhysRevLett.63.398.
[10] W.A. Crichton, M. Mezouar, T. Grande, S. Stølen, A. Grzechnik, Breakdown of intermediate-range order in liquid GeSe2 at high pressure, Nature 414 (2001) 622–625. https://doi.org/10.1038/414622a.
[11] P.F. McMillan, M. Wilson, D. Daisenberger, D. Machon, A density-driven phase transition between semiconducting and metallic polyamorphs of silicon, Nature Mater 4 (2005) 680–684. https://doi.org/10.1038/nmat1458.
[12] H.W. Sheng, H.Z. Liu, Y.Q. Cheng, J. Wen, P.L. Lee, W.K. Luo, S.D. Shastri, E. Ma, Polyamorphism in a metallic glass, Nature Mater 6 (2007) 192–197. https://doi.org/10.1038/nmat1839.
[13] Q. Zeng, Y. Ding, W.L. Mao, W. Yang, Stas.V. Sinogeikin, J. Shu, H. Mao, J.Z. Jiang, Origin of Pressure-Induced Polyamorphism in Ce 75 Al 25 Metallic Glass, Phys. Rev. Lett. 104 (2010) 105702. https://doi.org/10.1103/PhysRevLett.104.105702.
[14] S. Wei, F. Yang, J. Bednarcik, I. Kaban, O. Shuleshova, A. Meyer, R. Busch, Liquid–liquid transition in a strong bulk metallic glass-forming liquid, Nat Commun 4 (2013) 2083. https://doi.org/10.1038/ncomms3083.
[15] Q. Zeng, Z. Zeng, H. Lou, Y. Kono, B. Zhang, C. Kenney-Benson, C. Park, W.L. Mao, Pressure-induced elastic anomaly in a polyamorphous metallic glass, Applied Physics Letters 110 (2017) 221902. https://doi.org/10.1063/1.4984746.
[16] Q. Du, X.-J. Liu, Q. Zeng, H. Fan, H. Wang, Y. Wu, S.-W. Chen, Z.-P. Lu, Polyamorphic transition in a transition metal based metallic glass under high pressure, Phys. Rev. B 99 (2019) 014208. https://doi.org/10.1103/PhysRevB.99.014208.
[17] S. Ali Khan, X.-D. Wang, A. Saeed Ahmad, Q.-P. Cao, D.-X. Zhang, Y.-Z. Fang, H. Wang, J.-Z. Jiang, Temperature- and Pressure-Induced Polyamorphic Transitions in AuCuSi Alloy, J. Phys. Chem. C 123 (2019) 20342–20350. https://doi.org/10.1021/acs.jpcc.9b05167.
[18] Y. Cao, M. Yang, Q. Du, F.-K. Chiang, Y. Zhang, S.-W. Chen, Y. Ke, H. Lou, F. Zhang, Y. Wu, H. Wang, S. Jiang, X. Zhang, Q. Zeng, X. Liu, Z. Lu, Continuous polyamorphic transition in high-entropy metallic glass, Nat Commun 15 (2024) 6702. https://doi.org/10.1038/s41467-024-51080-8.
[19] Q. Luo, G. Garbarino, B. Sun, D. Fan, Y. Zhang, Z. Wang, Y. Sun, J. Jiao, X. Li, P. Li, N. Mattern, J. Eckert, J. Shen, Hierarchical densification and negative thermal expansion in Ce-based metallic glass under high pressure, Nat Commun 6 (2015) 5703. https://doi.org/10.1038/ncomms6703.

[20] W. Dmowski, G.H. Yoo, S. Gierlotka, H. Wang, Y. Yokoyama, E.S. Park, S. Stelmakh, T. Egami, High Pressure Quenched Glasses: unique structures and properties, Sci Rep 10 (2020) 9497. https://doi.org/10.1038/s41598-020-66418-7.

[21] O. Mishima, L.D. Calvert, E. Whalley, An apparently first-order transition between two amorphous phases of ice induced by pressure, Nature 314 (1985) 76–78. https://doi.org/10.1038/314076a0.

[22] I. Saika-Voivod, P.H. Poole, F. Sciortino, Fragile-to-strong transition and polyamorphism in the energy landscape of liquid silica, Nature 412 (2001) 514–517. https://doi.org/10.1038/35087524.

[23] S. Sen, S. Gaudio, B.G. Aitken, C.E. Lesher, Observation of a Pressure-Induced First-Order Polyamorphic Transition in a Chalcogenide Glass at Ambient Temperature, Phys. Rev. Lett. 97 (2006). https://doi.org/10.1103/physrevlett.97.025504.

[24] H. Luan, X. Zhang, H. Ding, F. Zhang, J.H. Luan, Z.B. Jiao, Y.-C. Yang, H. Bu, R. Wang, J. Gu, C. Shao, Q. Yu, Y. Shao, Q. Zeng, N. Chen, C.T. Liu, K.-F. Yao, High-entropy induced a glass-to-glass transition in a metallic glass, Nat Commun 13 (2022) 2183. https://doi.org/10.1038/s41467-022-29789-1.

[25] B. Ruta, V.M. Giordano, L. Erra, C. Liu, E. Pineda, Structural and dynamical properties of Mg65Cu25Y10 metallic glasses studied by in situ high energy X-ray diffraction and time resolved X-ray photon correlation spectroscopy, Journal of Alloys and Compounds 615 (2014) S45–S50. https://doi.org/10.1016/j.jallcom.2013.12.162.

[26] V.M. Giordano, B. Ruta, Unveiling the structural arrangements responsible for the atomic dynamics in metallic glasses during physical aging, Nat Commun 7 (2016) 10344. https://doi.org/10.1038/ncomms10344.

[27] S. Hechler, B. Ruta, M. Stolpe, E. Pineda, Z. Evenson, O. Gross, A. Bernasconi, R. Busch, I. Gallino, Microscopic evidence of the connection between liquid-liquid transition and dynamical crossover in an ultraviscous metallic glass former, Phys. Rev. Materials 2 (2018). https://doi.org/10.1103/physrevmaterials.2.085603.

[28] K.N. Pham, A.M. Puertas, J. Bergenholtz, S.U. Egelhaaf, A. Moussaïd, P.N. Pusey, A.B. Schofield, M.E. Cates, M. Fuchs, W.C.K. Poon, Multiple Glassy States in a Simple Model System, Science 296 (2002) 104–106. https://doi.org/10.1126/science.1068238.

[29] M. Frey, N. Neuber, S.S. Riegler, A. Cornet, Y. Chushkin, F. Zontone, L.M. Ruschel, B. Adam, M. Nabahat, F. Yang, J. Shen, F. Westermeier, M. Sprung, D. Cangialosi, V. Di Lisio, I. Gallino, R. Busch, B. Ruta, E. Pineda, Liquid-like versus stress-driven dynamics in a metallic glass former observed by temperature scanning X-ray photon correlation spectroscopy, Nat Commun 16 (2025) 4429. https://doi.org/10.1038/s41467-025-59767-2.

[30] A. Madsen, A. Fluerasu, B. Ruta, Structural Dynamics of Materials Probed by X-Ray Photon Correlation Spectroscopy, in: E. Jaeschke, S. Khan, J.R. Schneider, J.B. Hastings (Eds.), Synchrotron Light Sources and Free-Electron Lasers, Springer International Publishing, Cham, 2015: pp. 1–21. https://doi.org/10.1007/978-3-319-04507-8_29-1.

[31] A. Cornet, A. Ronca, J. Shen, F. Zontone, Y. Chushkin, M. Cammarata, G. Garbarino, M. Sprung, F. Westermeier, T. Deschamps, B. Ruta, High-pressure X-ray photon correlation spectroscopy at fourth-generation synchrotron sources, J Synchrotron Rad 31 (2024) 527–539. https://doi.org/10.1107/S1600577524001784.

[32] A. Karina, H. Li, T. Eklund, M. Ladd-Parada, B. Massani, M. Filianina, N. Kondedan, A. Rydh, K. Holl, R. Trevorah, S. Huotari, R.P.C. Bauer, C. Goy, N.N. Striker, F. Dallari, F. Westermeier, M. Sprung, F. Lehmkühler, K. Amann-Winkel, Multicomponent dynamics in amorphous ice studied using X-ray photon correlation spectroscopy at elevated pressure and cryogenic temperatures, Commun Chem 8 (2025) 82. https://doi.org/10.1038/s42004-025-01480-8.

[33] X. Zhang, H. Lou, B. Ruta, Y. Chushkin, F. Zontone, S. Li, D. Xu, T. Liang, Z. Zeng, H. Mao, Q. Zeng, Pressure-induced nonmonotonic cross-over of steady relaxation dynamics in a metallic glass, Proc. Natl. Acad. Sci. U.S.A. 120 (2023) e2302281120. https://doi.org/10.1073/pnas.2302281120.

[34] A. Cornet, G. Garbarino, F. Zontone, Y. Chushkin, J. Jacobs, E. Pineda, T. Deschamps, S. Li, A. Ronca, J. Shen, G. Morard, N. Neuber, M. Frey, R. Busch, I. Gallino, M. Mezouar, G. Vaughan, B. Ruta, Denser glasses relax faster: Enhanced atomic mobility and anomalous particle displacement under in-situ high pressure compression of metallic glasses, Acta Materialia 255 (2023) 119065. https://doi.org/10.1016/j.actamat.2023.119065.

[35] A.D. Phan, A. Zaccone, V.D. Lam, K. Wakabayashi, Theory of Pressure-Induced Rejuvenation and Strain Hardening in Metallic Glasses, Phys. Rev. Lett. 126 (2021) 025502. https://doi.org/10.1103/PhysRevLett.126.025502.

[36] N.K. Ngan, A.D. Phan, A. Zaccone, Impact of High Pressure on Reversible Structural Relaxation of Metallic Glass, Physica Rapid Research Ltrs 15 (2021) 2100235. https://doi.org/10.1002/pssr.202100235.

[37] J. Pan, Yu.P. Ivanov, W.H. Zhou, Y. Li, A.L. Greer, Strain-hardening and suppression of shear-banding in rejuvenated bulk metallic glass, Nature 578 (2020) 559–562. https://doi.org/10.1038/s41586-020-2016-3.

[38] I. Gallino, On the Fragility of Bulk Metallic Glass Forming Liquids, Entropy 19 (2017) 483. https://doi.org/10.3390/e19090483.

[39] I. Gallino, D. Cangialosi, Z. Evenson, L. Schmitt, S. Hechler, M. Stolpe, B. Ruta, Hierarchical aging pathways and reversible fragile-to-strong transition upon annealing of a metallic glass former, Acta Materialia 144 (2018) 400–410. https://doi.org/10.1016/j.actamat.2017.10.060.

[40] H. Lou, Z. Zeng, F. Zhang, S. Chen, P. Luo, X. Chen, Y. Ren, V.B. Prakapenka, C. Prescher, X. Zuo, T. Li, J. Wen, W.-H. Wang, H. Sheng, Q. Zeng, Two-way tuning of structural order in metallic glasses, Nat Commun 11 (2020) 314. https://doi.org/10.1038/s41467-019-14129-7.

[41] E.-Y. Chen, S.-X. Peng, L. Peng, M. Di Michiel, G.B.M. Vaughan, Y. Yu, H.-B. Yu, B. Ruta, S. Wei, L. Liu, Glass-forming ability correlated with the liquid-liquid transition in Pd42.5Ni42.5P15 alloy, Scripta Materialia 193 (2021) 117–121. https://doi.org/10.1016/j.scriptamat.2020.10.042.

[42] G. Ashiotis, A. Deschildre, Z. Nawaz, J.P. Wright, D. Karkoulis, F.E. Picca, J. Kieffer, The fast azimuthal integration Python library: *pyFAI*, J Appl Crystallogr 48 (2015) 510–519. https://doi.org/10.1107/S1600576715004306.

[43] J. Kieffer, S. Petitdemange, T. Vincent, Real-time diffraction computed tomography data reduction, J Synchrotron Rad 25 (2018) 612–617. https://doi.org/10.1107/S1600577518000607.

[44] S. Boccato, Y. Garino, G. Morard, B. Zhao, F. Xu, C. Sanloup, A. King, N. Guignot, A. Clark, G. Garbarino, M. Morand, D. Antonangeli, Amorpheus: a Python-based software for the treatment of X-ray scattering data of amorphous and liquid systems, High Pressure Research 42 (2022) 69–93. https://doi.org/10.1080/08957959.2022.2032032.

[45] J. Krogh-Moe, A method for converting experimental X-ray intensities to an absolute scale, Acta Cryst 9 (1956) 951–953. https://doi.org/10.1107/S0365110X56002655.

[46] C.J. Smithells, T.C. Totemeier, W.F. Gale, Smithells metals reference book, 8th ed. / edited by W.F. Gale, T.C. Totemeier, Elsevier Butterworth-Heinemann, Amsterdam Boston, 2004.

[47] Y. Chushkin, C. Caronna, A. Madsen, A novel event correlation scheme for X-ray photon correlation spectroscopy, J Appl Crystallogr 45 (2012) 807–813. https://doi.org/10.1107/S0021889812023321.

[48] A. Madsen, A. Fluerasu, B. Ruta, Structural Dynamics of Materials Probed by X-Ray Photon Correlation Spectroscopy, in: E. Jaeschke, S. Khan, J.R. Schneider, J.B. Hastings (Eds.), Synchrotron Light Sources and Free-Electron Lasers, Springer International Publishing, Cham, 2015: pp. 1–21. https://doi.org/10.1007/978-3-319-04507-8_29-1.

[49] A.R. Yavari, A.L. Moulec, A. Inoue, N. Nishiyama, N. Lupu, E. Matsubara, W.J. Botta, G. Vaughan, M.D. Michiel, Å. Kvick, Excess free volume in metallic glasses measured by X-ray diffraction, Acta Materialia 53 (2005) 1611–1619. https://doi.org/10.1016/j.actamat.2004.12.011.

[50] S. Scudino, M. Stoica, I. Kaban, K.G. Prashanth, G.B.M. Vaughan, J. Eckert, Length scale-dependent structural relaxation in Zr57.5Ti7.5Nb5Cu12.5Ni10Al7.5 metallic glass, Journal of Alloys and Compounds 639 (2015) 465–469. https://doi.org/10.1016/j.jallcom.2015.03.179.

[51] Z. Evenson, S.E. Naleway, S. Wei, O. Gross, J.J. Kruzic, I. Gallino, W. Possart, M. Stommel, R. Busch, β relaxation and low-temperature aging in a Au-based bulk metallic glass: From elastic properties to atomic-scale structure, Phys. Rev. B 89 (2014) 174204. https://doi.org/10.1103/PhysRevB.89.174204.

[52] S. Chen, D. Xu, X. Zhang, X. Chen, Y. Liu, T. Liang, Z. Yin, S. Jiang, K. Yang, J. Zeng, H. Lou, Z. Zeng, Q. Zeng, Reversible linear-compression behavior of free volume in a metallic glass, Phys. Rev. B 105 (2022) 144201. https://doi.org/10.1103/PhysRevB.105.144201.

[53] Q. Zeng, Y. Kono, Y. Lin, Z. Zeng, J. Wang, S.V. Sinogeikin, C. Park, Y. Meng, W. Yang, H.-K. Mao, W.L. Mao, Universal Fractional Noncubic Power Law for Density of Metallic Glasses, Phys. Rev. Lett. 112 (2014) 185502. https://doi.org/10.1103/PhysRevLett.112.185502.

[54] J. Ding, E. Ma, M. Asta, R.O. Ritchie, Second-Nearest-Neighbor Correlations from Connection of Atomic Packing Motifs in Metallic Glasses and Liquids, Sci Rep 5 (2015) 17429. https://doi.org/10.1038/srep17429.

[55] O. Gross, N. Neuber, A. Kuball, B. Bochtler, S. Hechler, M. Frey, R. Busch, Signatures of structural differences in Pt–P- and Pd–P-based bulk glass-forming liquids, Commun Phys 2 (2019) 83. https://doi.org/10.1038/s42005-019-0180-2.

[56] N. Neuber, M. Sadeghilaridjani, N. Ghodki, O. Gross, B. Adam, L. Ruschel, M. Frey, S. Muskeri, M. Blankenburg, I. Gallino, R. Busch, S. Mukherjee, Effect of composition and thermal history on deformation behavior and cluster connections in model bulk metallic glasses, Sci Rep 12 (2022) 17133. https://doi.org/10.1038/s41598-022-20938-6.

[57] B. Ruta, G. Baldi, G. Monaco, Y. Chushkin, Compressed correlation functions and fast aging dynamics in metallic glasses, The Journal of Chemical Physics 138 (2013) 054508. https://doi.org/10.1063/1.4790131.

[58] B. Ruta, E. Pineda, Z. Evenson, Relaxation processes and physical aging in metallic glasses, J. Phys.: Condens. Matter 29 (2017) 503002. https://doi.org/10.1088/1361-648X/aa9964.

[59] M. Frey, N. Neuber, S.S. Riegler, A. Cornet, Y. Chushkin, F. Zontone, L. Ruschel, B. Adam, M. Nabahat, F. Yang, J. Shen, F. Westermeier, M. Sprung, D. Cangialosi, V.D. Lisio, I. Gallino, R. Busch, B. Ruta, E. Pineda, On the interplay of liquid-like and stress-driven dynamics in a metallic glass former observed by temperature scanning XPCS, (2024). https://doi.org/10.48550/arXiv.2403.12306.

[60] F. Perakis, K. Amann-Winkel, F. Lehmkühler, M. Sprung, D. Mariedahl, J.A. Sellberg, H. Pathak, A. Späh, F. Cavalca, D. Schlesinger, A. Ricci, A. Jain, B. Massani, F. Aubree, C.J. Benmore, T. Loerting, G. Grübel, L.G.M. Pettersson, A. Nilsson, Diffusive dynamics during the high-to-low density transition in amorphous ice, Proc. Natl. Acad. Sci. U.S.A. 114 (2017) 8193–8198. https://doi.org/10.1073/pnas.1705303114.

[61] M. Ladd-Parada, H. Li, A. Karina, K.H. Kim, F. Perakis, M. Reiser, F. Dallari, N. Striker, M. Sprung, F. Westermeier, G. Grübel, A. Nilsson, F. Lehmkühler, K. Amann-Winkel, Using coherent X-rays to follow dynamics in amorphous ices, Environ. Sci.: Atmos. 2 (2022) 1314–1323. https://doi.org/10.1039/D2EA00052K.

[62] H. Li, M. Ladd-Parada, A. Karina, F. Dallari, M. Reiser, F. Perakis, N.N. Striker, M. Sprung, F. Westermeier, G. Grübel, W. Steffen, F. Lehmkühler, K. Amann-Winkel, Intrinsic Dynamics of Amorphous Ice Revealed by a Heterodyne Signal in X-ray Photon Correlation Spectroscopy Experiments, J. Phys. Chem. Lett. 14 (2023) 10999–11007. https://doi.org/10.1021/acs.jpclett.3c02470.

[63] L. Müller, M. Waldorf, C. Gutt, G. Grübel, A. Madsen, T.R. Finlayson, U. Klemradt, Slow Aging Dynamics and Avalanches in a Gold-Cadmium Alloy Investigated by X-Ray Photon Correlation Spectroscopy, Phys. Rev. Lett. 107 (2011). https://doi.org/10.1103/physrevlett.107.105701.

[64] C. Sanborn, K.F. Ludwig, M.C. Rogers, M. Sutton, Direct Measurement of Microstructural Avalanches during the Martensitic Transition of Cobalt Using Coherent X-Ray Scattering, Phys. Rev. Lett. 107 (2011). https://doi.org/10.1103/physrevlett.107.015702.

[65] R. Jeanloz, R.M. Hazen, Finite-strain analysis of relative compressibilities: Application to the high-pressure wadsleyite phase as an illustration, American Mineralogist 76 (1991) 1765–1768.

[66] W.H. Wang, The elastic properties, elastic models and elastic perspectives of metallic glasses, Progress in Materials Science 57 (2012) 487–656. https://doi.org/10.1016/j.pmatsci.2011.07.001.

[67] Z. Zhou, H. Wang, M. Li, Hydrostatic pressure effect on metallic glasses: A theoretical prediction, Journal of Applied Physics 126 (2019) 145901. https://doi.org/10.1063/1.5118221.

[68] H.B. Lou, Y.K. Fang, Q.S. Zeng, Y.H. Lu, X.D. Wang, Q.P. Cao, K. Yang, X.H. Yu, L. Zheng, Y.D. Zhao, W.S. Chu, T.D. Hu, Z.Y. Wu, R. Ahuja, J.Z. Jiang, Pressure-induced amorphous-to-amorphous configuration change in Ca-Al metallic glasses, Sci Rep 2 (2012) 376. https://doi.org/10.1038/srep00376.

[69] L. Zhang, F. Sun, X. Hong, J. Wang, G. Liu, L. Kong, H. Yang, X. Liu, Y. Zhao, W. Yang, Pressure-induced polyamorphism by quantitative structure factor and pair distribution function analysis in two Ce-based metallic glasses, Journal of Alloys and Compounds 695 (2017) 1180–1184. https://doi.org/10.1016/j.jallcom.2016.10.246.

# Supplementary Material:

# Emergence of multiple relaxation processes during low to high density transition in Au$_{49}$Cu$_{26.9}$Si$_{16.3}$Ag$_{5.5}$Pd$_{2.3}$ metallic glass

Alberto Ronca* *et al.*

*Corresponding authors. Emails: alberto.ronca@neel.cnrs.fr; beatrice.ruta@neel.cnrs.fr

1. **Structural changes during compression/decompression cycle**

The position of the second diffraction peak and its fit is reported in Fig. S1. As for the first sharp diffraction peak (FSDP) described in the main manuscript, the second peak of the $S(q)$ exhibits a reversible densification with no particular changes in the pressure dependence of the center of mass upon decompressing.

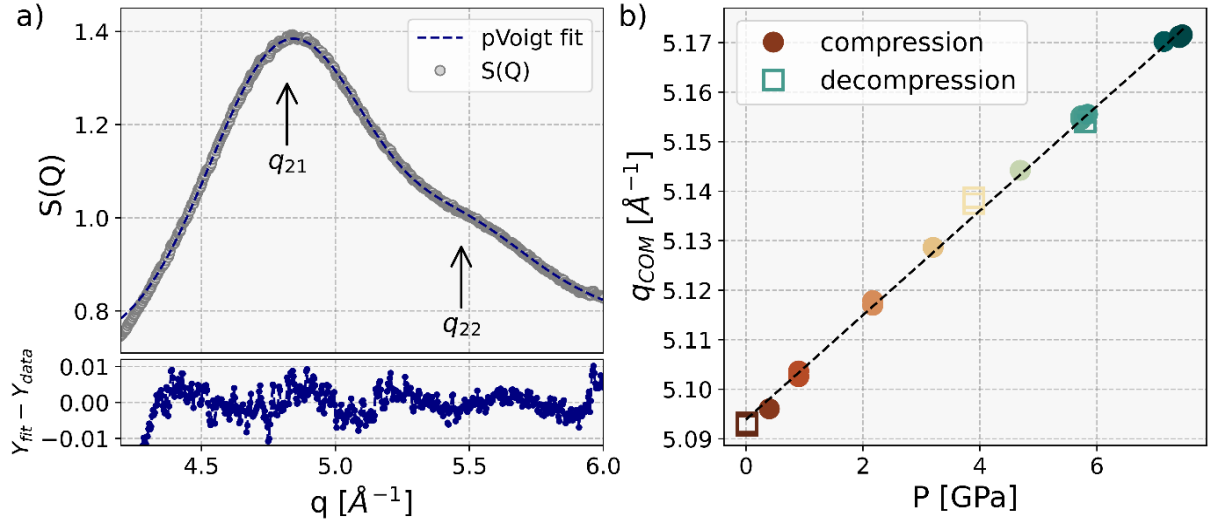

**Fig. S1. a)** Second diffraction peak of the $S(Q)$. The blue dotted line is the best fit of the experimental data with a sum of two Pseudo-Voigt function. $q_{21}$ and $q_{22}$ represents the position of the two sub-peaks. Fitting residuals are displayed in the plot on the bottom. **b)** Evolution of the center of mass $q_{COM}$ with pressure during both compression and decompression. $q_{COM}$ is estimated as $q_{COM} = (q_{21} * I_{21} + q_{22} * I_{22})/(I_{21} + I_{22})$, where $I_{21}$ and $I_{21}$ denotes the intensity of the respective sub-peak. The dotted line is a linear fit to the data, yielding a slope of $1.1 \times 10^{-2}$ Å$^{-1}$/GPa.

The position of the FSDP, $q_1$, can be used to estimate the average atomic volume changes of the system with pressure through $V/V_0 = (q_{1_0}/q_1)^D$, where $V_0$ and $q_{1_0}$ are the volume and the FSDP position at 1 atm, and $V(P)$ and $q_1(P)$ are the pressure dependent quantities. For macroscopically isotropic disorder systems, a cubic power law scaling ($D = 3$) is expected. However, a parameter $D = 2.5$ was found to reproduce experimental data in several MGs[1]. Fig. S2 shows the pressure induced volume changes estimated with the two power laws for completeness. Volumetric data can be fitted with the third-order Birch-Murnaghan isothermal equation of state (BM-EOS)[2] which yields an isothermal bulk modulus $B_0 = 153.6 \pm 4.7$ GPa and its pressure derivative $B_0' = 5.1 \pm 1.7$ for $D = 3$ and $B_0 = 184.3 \pm 5.7$ GPa, $B_0' = 6.1 \pm 2.1$ for $D = 2.5$. The bulk modulus obtained by fitting is consistent within its error bar with previously tabulated values[3].

The relative density increase occurring during compression can be directly estimated as $V_0/V - 1$, accounting for a 4.4 and 3.7 % increase in case of $D = 3$ and 2.5, respectively. A first estimation of the compressibility of the material and in particular its pressure dependence (Fig. 5 of the main text) can be obtained through the numerical calculation of the pressure derivative of the volume. The compressibility is then calculated following its definition as $k = -1/V \frac{\partial V}{\partial P}$, or, in our case, $k = -V_0/V \frac{\partial V/V_0}{\partial P}$. An analogous way of computing the compressibility is through the $(P - V)$ equation of state, expressed in terms of the Eulerian strain $f$ and the Normalized stress $F$[4], where $f = [(V/V_0)^{-2/3} - 1]/2$ and $F = P/[3f(1 + 2f)^{2.5}]$, respectively. The equation of state then takes the form $F = K(1 + af + \cdots)$, and in case of a linear relation for $F(f)$ it yields the bulk modulus $K$, and thus the compressibility, $k = 1/K$, as the intercept of the fit. The dependence $F(f)$, reported in Fig. S2 (right), exhibits two linear regimes, indicating a sudden change in the system compressibility at $\approx 3$ GPa, from a value of $(7.2 \pm 0.1)$ GPa$^{-1}$ to $(6.3 \pm 0.1)$ GPa$^{-1}$, confirming the existence of a pressure-induced crossover in compressibility, as showed also in Fig.5 of the main text with a different method.

The evolution of additional structural parameters with pressure is reported in Fig. S3 for the FWHM, and Fig. S4 for the intensity and area of the FSDP. Table 1 reports instead the main atomic pairs contributing to the $S(Q)$ and $G(r)$ functions, and Fig. S5 and S6 details on the G(r) functions.

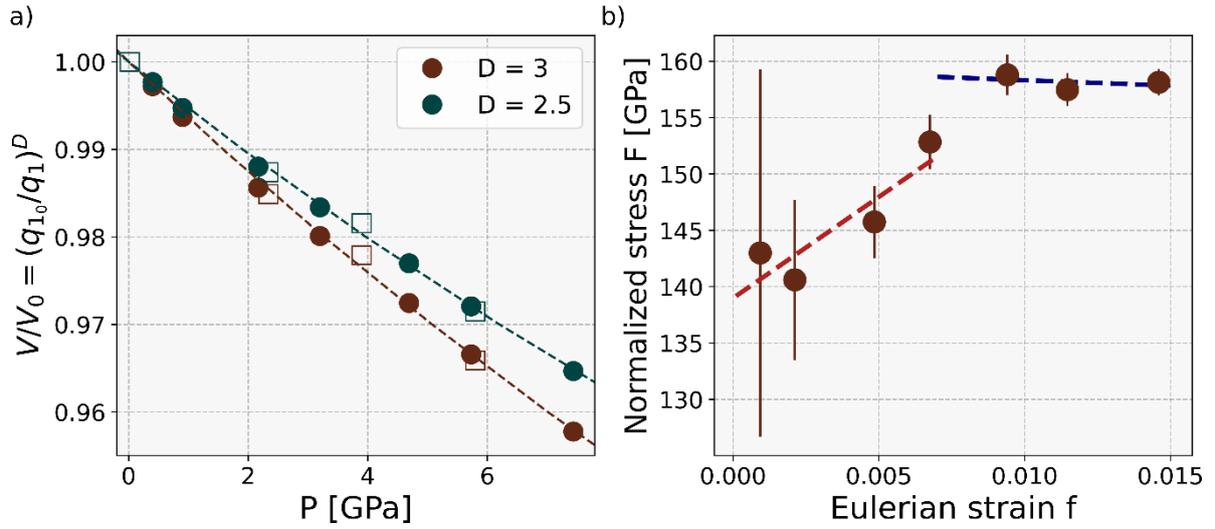

**Fig. S2. a)** Relative atomic volume changes during compression (full symbols) and decompression (empty symbols) obtained as $V/V_0 = (q_{1_0}/q_1)^D$, with $D = 2.5$ and 3. The dotted lines are the best fit of the third-order Birch-Murnaghan isothermal equation of state (BM-EOS) to the data. **b)** Normalized stress $F$ as a function of the Eulerian strain $f$. Dotted lines are linear fits in the two identified regimes, below and above $\approx 3$ GPa, respectively. In the case of a linear dependence of F in f, the intercept of the fit yield the bulk modulus of the system.

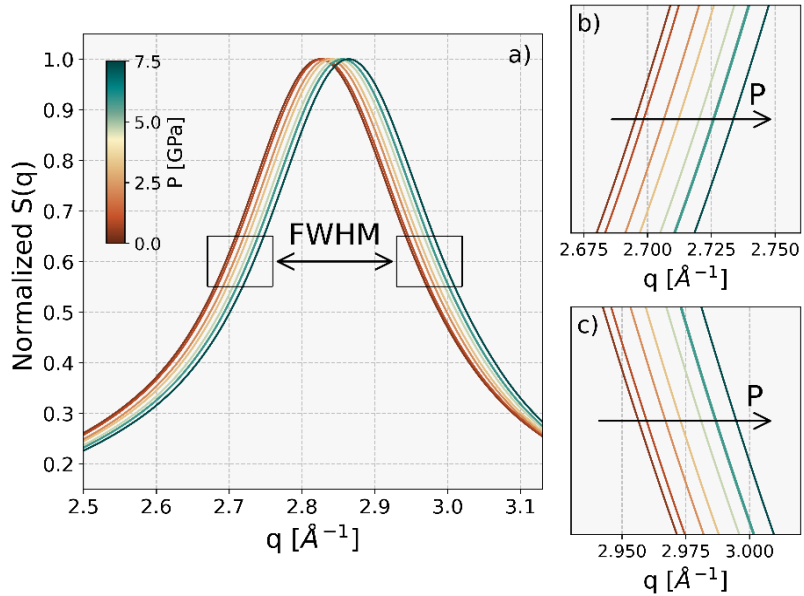

**Fig. S3. a)** Normalized FSDP as a function of pressure. Magnified left **(b)** and right **(c)** part of the FSDP at half maximum.

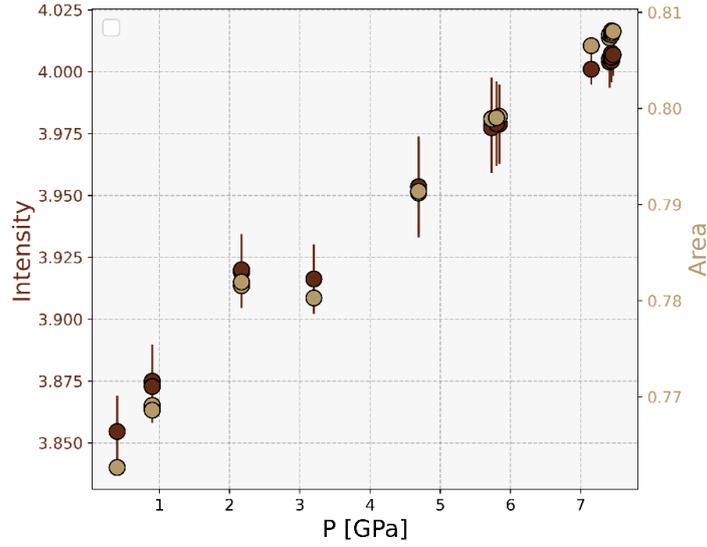

**Fig. S4.** Comparison of FSDP intensity and area evolution with pressure. The good match between the two quantities confirms the observed phenomenology.

| Atomic pair | Bond length [Å] | $w_{ij}$ |
|---|---|---|
| **Au-Au** | 2.88 | 0.57 |
| **Au-Cu** | 2.72 | 0.23 |
| **Au-Ag** | 2.88 | 0.07 |
| **Au-Si** | 2.55 | 0.04 |
| **Cu-Cu** | 2.56 | 0.02 |

**Table S1.** Main atomic pairs contributing to the $S(Q)$ and $G(r)$ function of the $Au_{49}Cu_{26.9}Si_{16.3}Ag_{5.5}Pd_{2.3}$ MG. The ambient conditions Faber-Ziman[5] weighting factors $w_{ij}$ [6] are shown together with the bond lengths, estimated using Goldschmidt metallic radii, which reflect the typical atomic sizes in close-packed structure, as the ones of metallic glasses. The covalent radius is used in the case of Silicon.

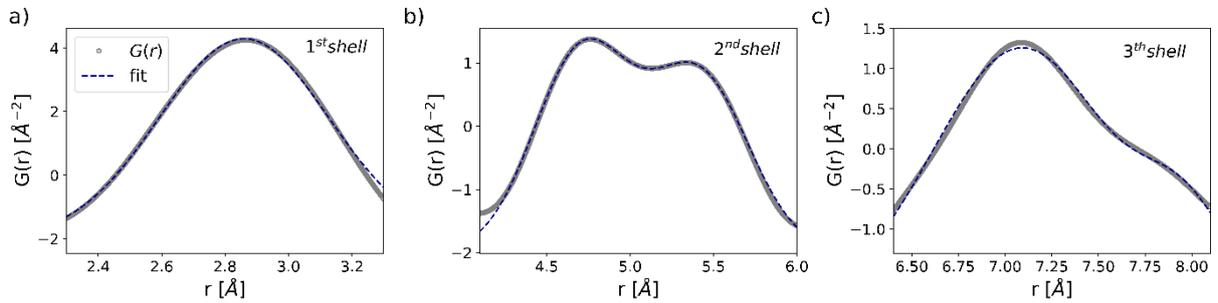

**Fig. S5.** First **(a)**, second **(b)** and third **(c)** shell of the reduced pair distribution function $G(r)$. Blue dotted lines are the best fit of experimental data with a single and a sum of two Gaussian functions, respectively.

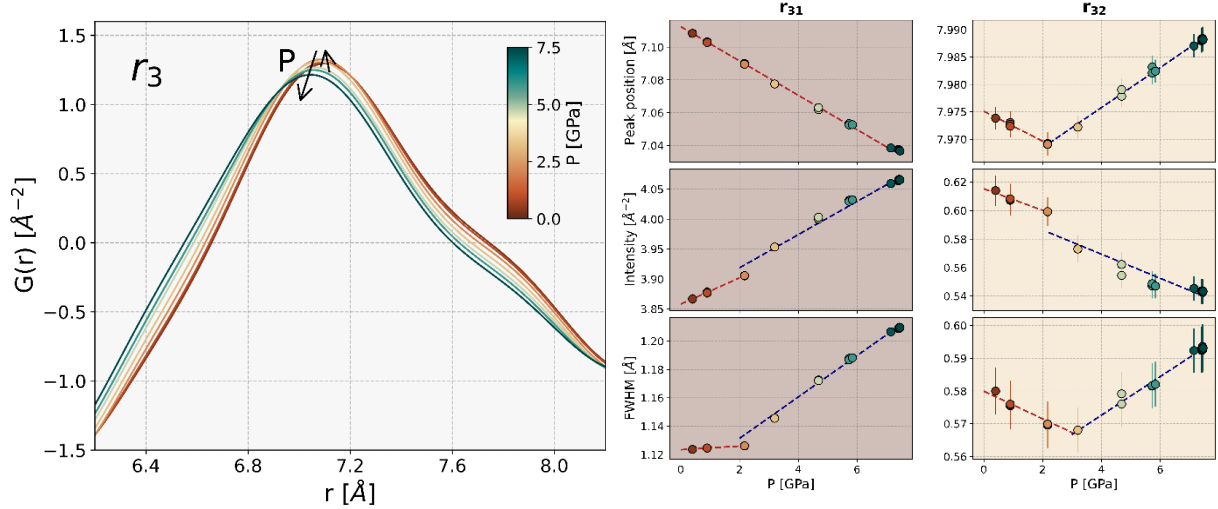

**Fig. S6. (left)** Pressure evolution of the 3$^{th}$ diffraction shell of the reduced pair distribution function $G(r)$. The arrow indicates the pressure behaviors of the peak. **(right)** Pressure dependence of the respective parameters obtained by fitting the 3$^{th}$ shell with a sum of two Gaussian functions ($r_{31}$ and $r_{32}$). Peak positions (**top row**), intensities (**middle row**) and FWHM (**bottom row**). Dotted lines are linear fit of the data. A single linear fit fails to capture the pressure evolution of the second peak position $r_{32}$. All the other fitting parameters exhibits a more or less evident change in the 2 – 3 GPa pressure range, and are modelled accordingly (red and blue dotted lines).

## 2. Pressure dependence of the atomic dynamics

Fig. S7 shows the pressure protocol that we have followed during the XPCS measurements. The starting pressure of 0.4 GPa corresponds to the loading pressure of the DAC. Data taken in these conditions are discarded in order to have the same history at all pressures. Multiple XPCS scans were taken during each isobar. Upon decompression, the dynamics slow down considerably compared to compression, making measurements of relaxation decays challenging on comparable timescales. This dynamical hysteresis was already observed in a Pt-based MG[7], and will be the subject of future works.

The dynamical crossover at 3.2 GPa is shown in Fig. S8. For pressure beyond the 3.2 GPa threshold, the activation of a second relaxation process superimposed to the predominant compressed decorrelation component is well evident in Fig. S8b) and d) by the secondary decorrelation component in the $g_2(q,t)$ functions and by the broad green-yellow region surrounding the high-intensity diagonal of the two-times correlation function (TTCF), which represents a time resolved version of the ISF. The secondary decay in the $g_2(q,t)$ occurs at almost one order of magnitude longer time-scales and with reduced strength with respect the predominant one. This behaviour is reproducible when probing a fresh sample region. To exclude local beam-induced effects, XPCS data were collected ≈ 30 μm away from the original spot – well beyond the ≈ 5x4 $μm^2$ x-rays beam size – ensuring no prior irradiation. As shown in Fig. S9, the dynamics at both positions exhibit a similar double relaxation decay, suggesting that this behaviour reflects a global feature of the system rather than a localized, possibly related to beam induced effects.

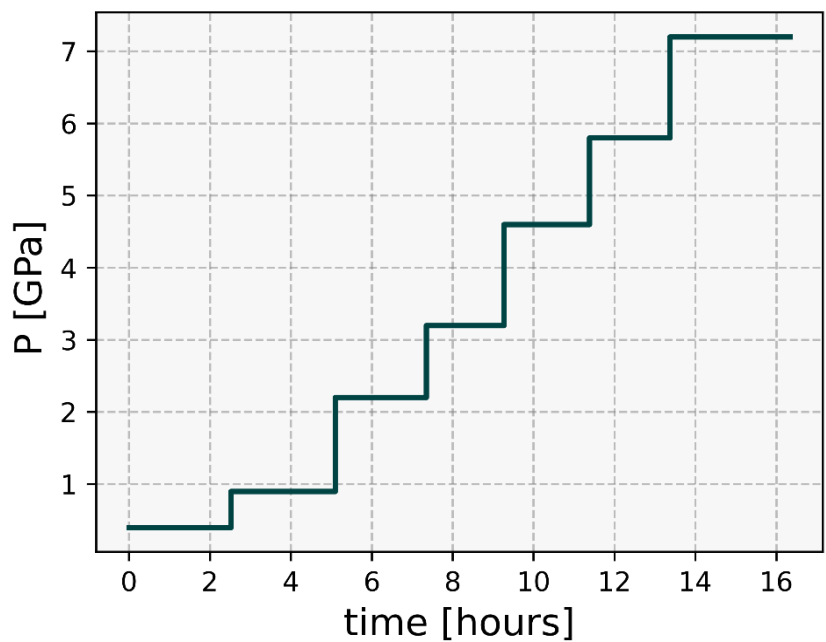

**Fig. S7. Pressure protocol of the XPCS measurements.** Multiple XPCS scans were taken during each isobar.

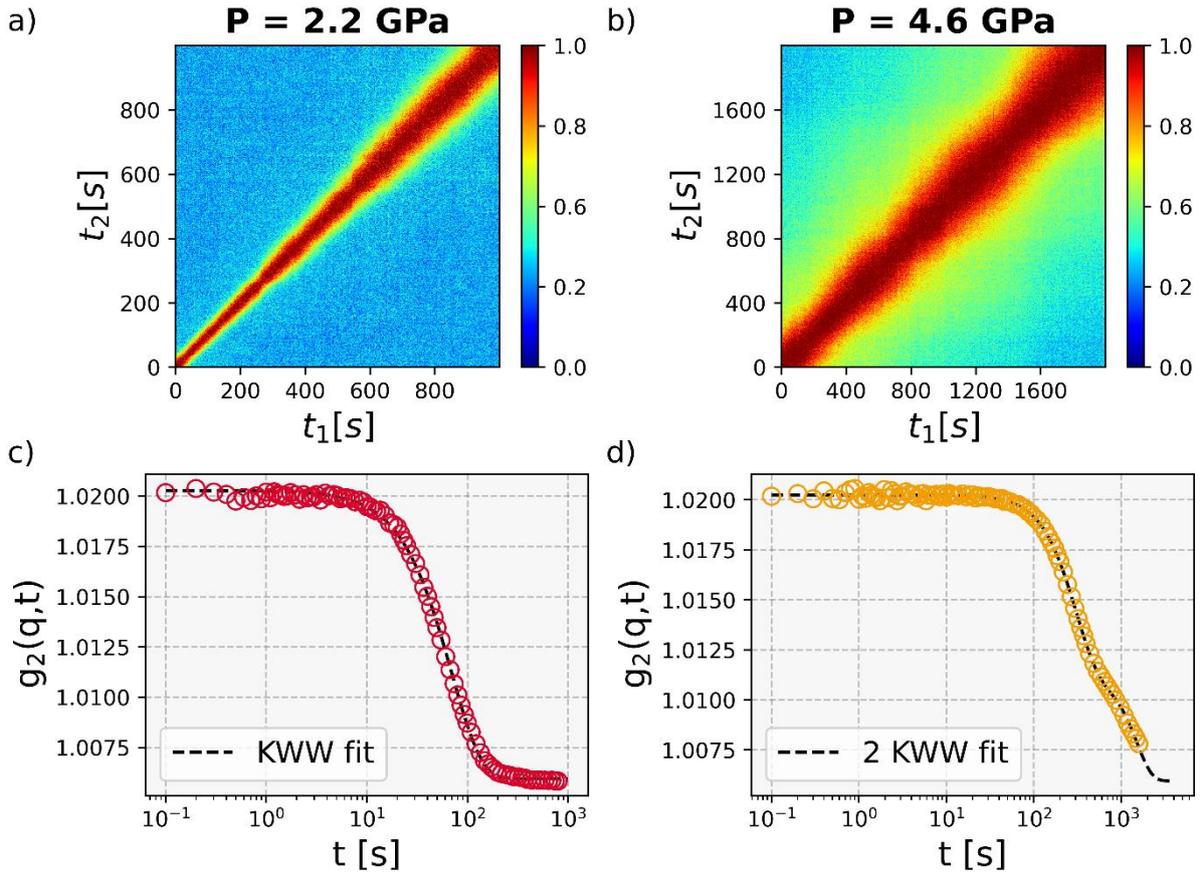

**Fig. S8. Emergence of a second relaxation process after 3.2 GPa.** Example of TTCFs measured before the transition at 2.2 GPa **(a)** and after the transition at 4.6 GPa **(b)**. Corresponding $g_2(q,t)$ functions are shown in panels **c)** and **d)**. A second relaxation process which was not shown at previous pressures emerges at pressures larger than 3.2 GPa, as indicated by the additional decay appearing in the $g_2(q,t)$ in panel d) and from the broad green-yellowish region surrounding that high-intensity red area of the TTCFs in panel b).

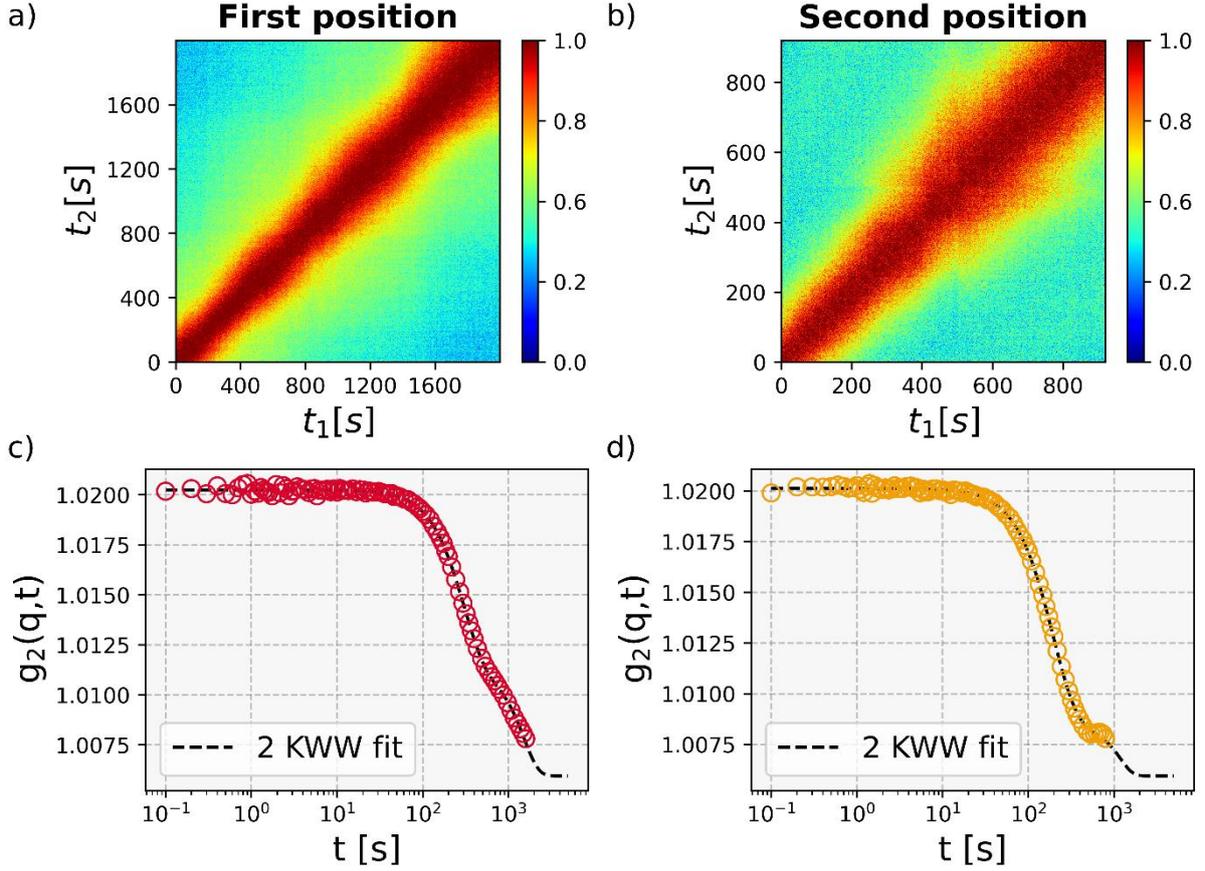

**Fig. S9. Double relaxation dynamics at distinct sample positions. (a – b)** TTCFs acquired at 4.6 GPa from two sample positions ≈ 30 $\mu$m apart. **(c – d)** Corresponding $g_2(q,t)$ functions. The second relaxation process in the second sample position can be seen by both the broad green-yellowish area surrounding that high-intensity diagonal of the TTCF and by the second decay of the $g_2(q,t)$, which is visible despite full decorrelation is not reached.

Figure S10 shows the results of using two distinct model functions, consisting of the sum of two Kohlrausch Williams Watt (KWW) terms, to describe the ISFs showing a two-step decay (Fig. 4 in the main manuscript):

$$g_2(q,t) - 1 = c \left[ a \exp\left(-\left(\frac{t}{\tau_1}\right)^{\beta_1}\right) + (a-1) \exp\left(-\left(\frac{t}{\tau_2}\right)^{\beta_2}\right) \right]^2 \quad (1)$$

$$g_2(q,t) - 1 = c_1 \exp\left(-2\left(\frac{t}{\tau_1}\right)^{\beta_1}\right) + c_2 \exp\left(-2\left(\frac{t}{\tau_2}\right)^{\beta_2}\right) \quad (2)$$

The first equation includes a mixed term. The two models reproduce comparable $\tau_1$ and $\tau_2$, $\tau_1$ being the relaxation time of the dominant relaxation process at shorted time scales and with greater strength than the second process ($\tau_2$). Both models predict compressed-shaped KWW for both processes.

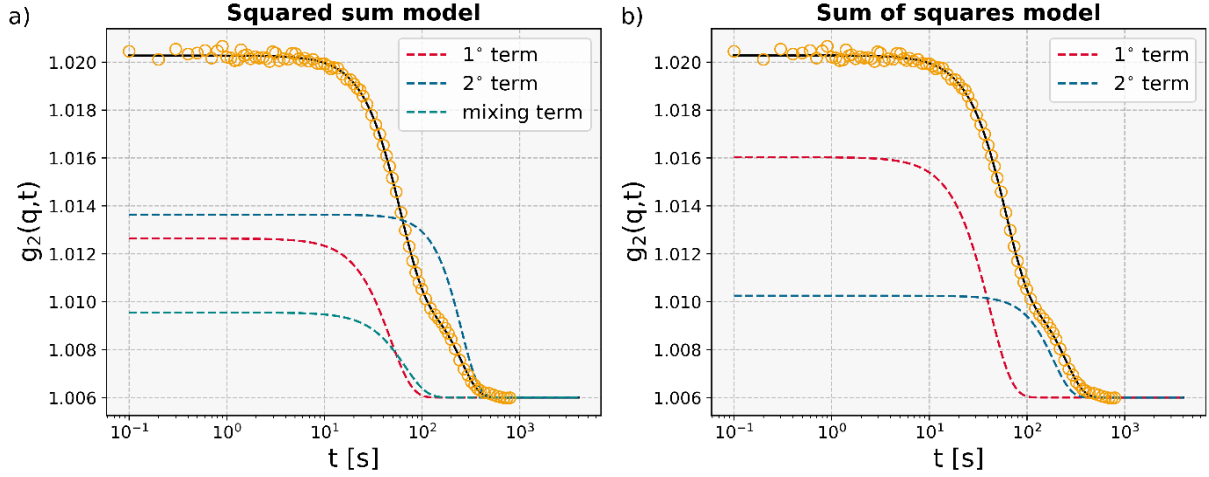

**Fig. S10. Comparison of different additive KWW model functions.** Fits of the experimental data with equation (1) **(a)** and (2) **(b)**, showed as black dotted lines. The different terms of each model are displayed as colored dotted lines together with the experimental data (yellow empty circles) which have been measured at 5.8 GPa.

3. **Observation on the dynamical transition occurring at 3.2 GPa and the emergence of a second relaxation decay at high pressure**

The avalanche-like dynamics observed at 3.2 GPa cannot be attributed to experimental artefacts. Pressure was stable throughout the whole XPCS measurements within the uncertainty of 0.1 GPa[8] at all measured pressures, ruling out mechanical instabilities within the DAC as a source of the detected dynamics at the probed pressure. The pressure is changed remotely from another room dismissing also the possibility of any human accident at the sample stage. Furthermore, the absence of a characteristic frequency for the avalanches whose occurrence decreases with increasing physical aging dismisses also the possible association with a mechanical or electronic vibrations (whose frequency would be constant and anyway orders of magnitude faster).

We rule out also possible contributions coming from the pressure transmitting medium (PTM) in the DAC. The choice of nitrogen as PTM allows for hydrostatic conditions in the 0 – 10 GPa range, comparably to those obtained with helium and neon[9]. Solid Nitrogen is expected to show a phase transition between the orientationally disordered $\beta$ and $\delta$ phases around 4.6 GPa[10]. By means of Raman Spectroscopy we have followed the aforementioned transition to check i) the possible existence of a change in the PTM at 3.2 GPa, and ii) the temporal nature of the transition. The β and δ phases are manifested by the splitting of the Raman peak, as showed in Fig. S11a, where the β-$N_2$ and δ-$N_2$ Raman spectra measured at pressure below and above 4.6 GPa are displayed, respectively. To compare with XPCS data we have collected Raman spectra immediately below 4.6 GPa for more than 3 hours as displayed in Fig. S11b, indicating the static nature of the transition. The pressure evolution of the Raman peaks is presented in Fig. 11c, even upon decompression. Overall, the β- δ transition appears at 4.8 GPa, with a reversible nature and no hysteresis upon decompression and without any kinetic effects. All these arguments indicate that the avalanches dynamics observed at 3.2 GPa and the consequent emergence of a second relaxation process are indeed strictly related to investigated metallic sample and cannot be attributed to any external artefacts.

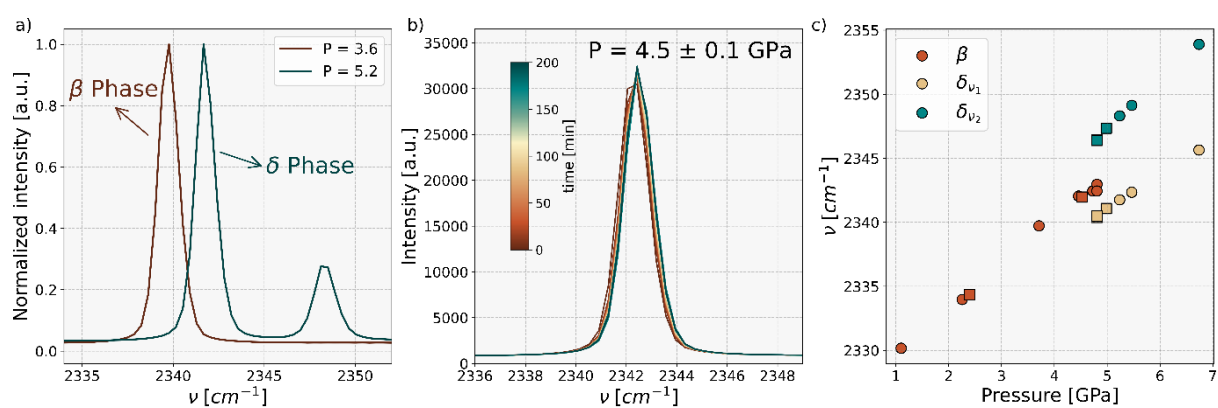

**Fig. S11. Vibrational Raman shift of solid nitrogen. (a)** Raman spectra of β-$N_2$ and δ-$N_2$ measured at ambient temperature and at different pressures. A phase transition between the β- and δ- phases is expected around 4.6 GPa. **(b)** Successive Raman spectra of the β-$N_2$ measured at 4.5 GPa for more than 3 hours. No kinetic effects are observed. **(c)** Room temperature Raman shifts for β-$N_2$, having only one Raman peak β, and δ-$N_2$, with two Raman peaks $\nu_1$ and $\nu_2$. Data acquired during compression and decompression are displayed as full circles and squares, respectively. The completely reversible transition with no hysteresis is showed in the data.

**References**

[1] Q. Zeng, Y. Kono, Y. Lin, Z. Zeng, J. Wang, S.V. Sinogeikin, C. Park, Y. Meng, W. Yang, H.-K. Mao, W.L. Mao, Universal Fractional Noncubic Power Law for Density of Metallic Glasses, Phys. Rev. Lett. 112 (2014) 185502. https://doi.org/10.1103/PhysRevLett.112.185502.
[2] F. Birch, Elasticity and constitution of the Earth's interior, J. Geophys. Res. 57 (1952) 227–286. https://doi.org/10.1029/JZ057i002p00227.
[3] W.H. Wang, The elastic properties, elastic models and elastic perspectives of metallic glasses, Progress in Materials Science 57 (2012) 487–656. https://doi.org/10.1016/j.pmatsci.2011.07.001.

[4] R. Jeanloz, R.M. Hazen, Finite-strain analysis of relative compressibilities: Application to the high-pressure wadsleyite phase as an illustration, American Mineralogist 76 (1991) 1765–1768.

[5] T.E. Faber, J.M. Ziman, A theory of the electrical properties of liquid metals: III. the resistivity of binary alloys, Philosophical Magazine 11 (1965) 153–173. https://doi.org/10.1080/14786436508211931.

[6] C.J. Smithells, T.C. Totemeier, W.F. Gale, Smithells metals reference book, 8th ed. / edited by W.F. Gale, T.C. Totemeier, Elsevier Butterworth-Heinemann, Amsterdam Boston, 2004.

[7] A. Cornet, Denser glasses relax faster: Enhanced atomic mobility and anomalous particle displacement under in-situ high pressure compression of metallic glasses, Acta Mat. 255 (2023). https://doi.org/10.1016/j.actamat.2023.119065.

[8] A. Cornet, A. Ronca, J. Shen, F. Zontone, Y. Chushkin, M. Cammarata, G. Garbarino, M. Sprung, F. Westermeier, T. Deschamps, B. Ruta, High-pressure X-ray photon correlation spectroscopy at fourth-generation synchrotron sources, J Synchrotron Rad 31 (2024) 527–539. https://doi.org/10.1107/S1600577524001784.

[9] S. Klotz, J.-C. Chervin, P. Munsch, G. Le Marchand, Hydrostatic limits of 11 pressure transmitting media, J. Phys. D: Appl. Phys. 42 (2009) 075413. https://doi.org/10.1088/0022-3727/42/7/075413.

[10] M.I.M. Scheerboom, J.A. Schouten, Orientational behavior of solid nitrogen at high pressures investigated by vibrational Raman spectroscopy, The Journal of Chemical Physics 105 (1996) 2553–2560. https://doi.org/10.1063/1.472121.